\documentclass[prl,twocolumn,epsfig,longeq]{revtex4}

\usepackage{graphicx}
\usepackage{dcolumn}
\usepackage{bm}
\usepackage{color}
\usepackage{subfig}
\usepackage{amsmath,amsfonts,amsthm,amssymb}
\usepackage{physics}
\usepackage{float}
\usepackage{amssymb,wasysym}
\usepackage[normalem]{ulem}

\newcommand{\NT}[1]{\textcolor{black}{#1}}

\begin{document}

\title{Order and density fluctuations near the boundary in sheared dense suspensions}
\date{November 2, 2022}
\author{Joia M. Miller}
\author{Daniel L. Blair}
\author{Jeffrey S. Urbach}
\email{urbachj@georgetown.edu}
\affiliation{
Department of Physics and Institute for Soft Matter Synthesis and Metrology,\\ Georgetown University, Washington, DC 20057.}






\begin{abstract}
We introduce a novel approach to reveal ordering fluctuations in sheared dense suspensions, 
using line scanning in a combined rheometer and laser scanning confocal microscope.  We validate the technique with a moderately dense suspension, observing modest shear-induced ordering and a nearly linear flow profile.  At high concentration ($\phi = 0.55$) and applied stress just below shear thickening, we report ordering fluctuations with high temporal resolution, and directly measure a decrease in order with distance from the suspension's bottom boundary as well as a direct correlation between order and particle concentration.  Higher applied stress produces shear thickening with large fluctuations in boundary stress which we find are accompanied by dramatic fluctuations in suspension flow speeds.  The peak flow rates are independent of distance from the suspension boundary, indicating that they 
likely
arise from transient jamming that creates solid-like aggregates of particles moving together, but only briefly because the high speed fluctuations are interspersed with regions flowing much more slowly, suggesting that shear thickening suspensions possess complex internal structural dynamics, even in relatively simple geometries.
\end{abstract}

\maketitle

\section{Introduction}
\label{intro}
Flowing dense suspensions of colloidal particles appear in a wide range of important industrial processes, and the presence of flow can dramatically modify the suspension microstructure, which in turn impacts flow properties (reviewed in  \cite{Vermant:2005tb,Morris:2009wo, Lettinga:2016wf}). In particular, the presence of simple shear modifies the well-understood phase behavior of dense suspensions of nearly monodisperse colloidal particles in equilibrium.  Modest shear promotes layering which enhances crystallization, while higher shear rates often disrupt crystalline order \cite{Chen:1992ur, Vermant:2005tb, Holmqvist:2005aa, Wu:2009vi,Derks:2009tc, Richard:2015tz,  Lettinga:2016wf}.
In addition, the bulk response of the suspension can be very sensitive to the material close to the confining boundaries \cite{Shereda:2010te, Cheng:2012tt, Xu:2013wt, Mackay:2014ua}.  Planar boundaries promote the formation of layers of particles parallel to the boundaries \cite{Shereda:2010te,villada:2022aa}, enabling more efficient shearing as layers slide past each other with reduced close particle interactions, and a resulting reduction in the local viscosity \cite{Kulkarni:2009un,Pieper:2016vo, Kucuksonmez:2020vl}.  The layering also enhances ordering within the layer, and in many circumstances the sliding layers show a high degree of hexagonal ordering, similar to what is seen in two dimensional colloidal crystals \cite{Myung:2013uf, Mackay:2014ua}. 
In many dense suspensions, the shear thinning arising from increased layering is followed by dramatic shear thickening when the applied stress exceeds a critical value, with the increase in viscosity attributable to a transition from primarily hydrodynamic particle interactions at low stress to frictional interactions at high stress (reviewed in \cite{Morris:2020uy}). The onset of frictional interactions will likely disrupt layering \cite{Lin:2018aa}, a scenario confirmed by recent computer simulations \cite{Goyal:2022wa}.  A variety of evidence suggests that strong shear thickening is accompanied by complex spatiotemporal dynamics, including fluctuations in flow speed, local stress, and particle concentration \cite{Boersma:1991wi, Lootens:2003ur,  Nakanishi:2012ux, Guy:2015wc, Nagahiro:2016wu, Hermes:2016wj, Rathee:2017ut, Saint-Michel:2018vw, Ovarlez:2020ul, Gauthier:2021vg, Maharjan:2021ud, Rathee:2020un, Rathee:2020wi, Ganapathy:2020aa, Rathee:2022vy}. 
These results highlight the need for methods that can probe the dynamics of dense colloidal suspension with high spatial and temporal resolution.  Here we introduce a new approach to assessing dynamic ordering using a combined rheometer and laser scanning confocal microscope \cite{Dutta:2013aa}, where the laser scanning is limited to one spatial direction. In this case the scan is perpendicular to the flow direction, and the suspension flow is primarily responsible for the temporal evolution of the recorded intensity.  We validate the technique using a moderately dense suspension (volume fraction $\phi = 0.52$), where only modest shear-induced ordering is observed.  At higher concentration ($\phi = 0.55$) and applied stress just below the onset of shear thickening, we measure ordering fluctuations with high temporal resolution, and directly measure the decrease in ordering with distance from the bottom boundary of the suspension. We also observe a direct correlation between ordering and particle concentration, with local regions of high order corresponding to high particle concentrations, consistent with recent observations from computer simulations \cite{Goyal:2022wa}.  Higher applied stress produces shear thickening, and we find that the large fluctuations in boundary stress that underlie thickening are accompanied by dramatic fluctuations in suspension flow rates.  The peak flow rates are independent of distance from the suspension boundary, indicating that they arise from transient jamming that creates solid like aggregates of particles moving together. Such aggregates must be short-lived because the high speed fluctuations are interspersed with regions flowing much more slowly, suggesting that shear thickening suspensions possess very complex internal structural dynamics, even in relatively simple geometries.

\section{Approach:} 
For particulate suspensions, particle velocities are often extracted from image sequences by particle tracking or correlation analysis (PIV) \cite{Poon:2009aa}.  Particle tracking typically requires that the time between images is small compared to the time for a particle to move an interparticle separation (so that each particle can be unambiguously identified in successive images).  Thus for a frame rate $f$ and an interparticle spacing $\delta$, the flow speed must satisfy $v_{max}^{imaging}<<f\delta$. The typical small $\delta$ in dense suspensions requires either a vanishingly small $v_{max}$ or extremely high frame rate $f$ for successful tracking. PIV can, in principle, handle larger displacements between images, but only if the particle configuration maintains a consistent distinctive pattern so that there is a clear unique peak in the spatial cross-correlation between images, which is typically not the case for the dense suspensions studied here.

\begin{figure*}
\includegraphics[width=1\textwidth]{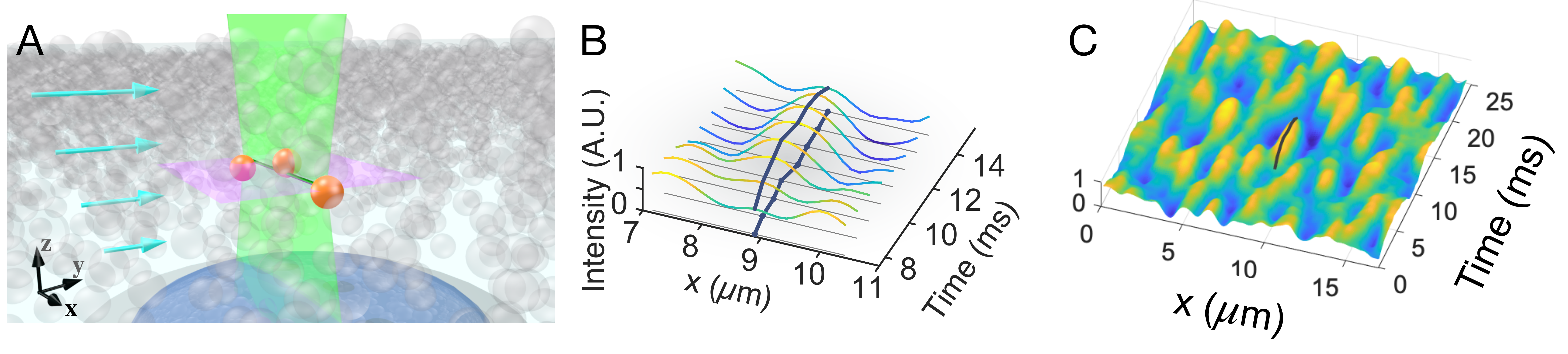}
\caption{ A) Schematic of linescan imaging of a sheared dense suspension.  The illumination laser rapidly scans back and forth, focused along the green line.  Particles transiting the line are illuminated (orange).  The imaging region for the usual 2D laser scan is indicated by the dashed line, and the arrows represent the average velocity field of the sheared suspension. 
The depth of focus (effective width of the image plane in $z$) depends on the optical setup, and is $\sim 1 \mu m$ here)
B) Graphical representation of the linescan data, where the continuous colored lines represent a small portion of each  individual scan, separated by a short interval.   A fluorescent particle transiting the scan region produces a local intensity maximum (marked with a black line).  C) Rendering of linescan data showing the passage of many particles.  The transit time can be determined by the length of the ridge (black line). (Note that in what follows the solvent is fluorescent but the particles are not, so the transit of particles produces local intensity minima.)}
\label{setup}
\end{figure*}

We use an alternative approach here, inspired by fluid dynamics and machine vision applications where the motion of material past  point or line detectors is used to infer speeds and structure \cite{Taylor:1938ul,He:2017ug}.   We take advantage of the rapid scan rate of a laser scanning confocal microscope, where the point of focus is rapidly scanned back and forth in one direction, which we will call the $x$ axis, with the normal rastering perpendicular to the scan line in the focal plane disabled  (Fig. \ref{setup}A). 
We illustrate the principle with a simple example:  A fluorescent particle transiting the line will produce a local intensity extremum, from which its $x$ position can be determined.  The time it takes the particle to transit the line can be measured by the length of the ridge produced when a position-time intensity surface is generated (Fig. \ref{setup}B).   If the center of the particle is in the plane of focus and neglecting the effects of the finite optical resolution,  the magnitude of the component of the particle velocity perpendicular to the line can be determined as $v=D/T$, where $D$ is the particle diameter and $T$ the transit time.
The requirement that the transit time is long compared to the time between scans means that the flow speed must be  $v_{max}^{line}<<sD$, where $s$ is the scan rate and $D$ the particle diameter. Because the scan rate for laser scanning confocal microscopy is 2-3 orders of magnitude larger than the frame rate (the  time to generate a 2D image is approximately the scan duration times the number of lines in an image), and the interparticle spacing in dense suspensions is comparable to the particle diameter, $\delta \sim D$, $v_{max}^{line}>>v_{max}^{imaging}$.  Figure \ref{setup}C shows a representative position-time intensity surface generate by our system (Leica SP5, scan rate 8000 Hz), where the passage of individual 1 $\mu$m diameter particles with transit times $\sim 5$ ms indicating speeds of $\sim 200 \mu m/s$ can be clearly identified despite the high particle density $\phi \approx 0.55$.  In practice individual particles cannot be robustly identified at from the data, and the position of the particles relative to the plane of focus is variable, so we instead extract speeds from characteristic lengths extracted from correlation analyses, as described below and detailed in Methods. The technique works equally well for non-fluorescent particles in a fluorescent suspending fluid, which is the approach we use in what follows.

\NT{The situation sketched above describes motion advection purely in the direction of the shear flow, but the linescan measurement will also be affected by other components of the flow velocity. Because of the geometry, the speed in the flow direction will normally be of order $\dot\gamma z$, where $\dot\gamma$ is the shear rate and  $z$ is the distance in the gradient direction from the bottom surface to the focus of the scan line (Fig. \ref{setup}A). In the cases considered here, the transverse velocity components are expected to be much smaller, but it is important to keep this consideration in mind when interpreting the linescan data.}

\NT{A related approach has been used to measure flow profiles in capillary and microfluidic systems (reviewed in \cite{koynov:212aa, Dong:2014aa}) where fluorescence correlation spectroscopy (FCS) is employed to extract speeds using fluorescent single molecules or nanoparticles \cite{kunst:2002aa}. The exact shape of the intensity autocorrelation function depends on the relative importance of diffusion and advection, but when diffusion is negligible compared to advection (as is the case in the situations considered here), the intensity autocorrelation function is a Gaussian with a width determined by the flow speed and the size of the illumination volume \cite{Gosch:2000aa}. By varying the scan orientation relative to the flow direction, the full (average) three dimensional velocity vector can be determined \cite{pan:2007aa, pan:2009aa}. In those applications, however, the volume fraction of the fluorescent species is always low, so they are reasonably assumed to be distributed randomly.  In the results presented below, spatial ordering dramatically affects the spatiotemporal intensity fluctuations, complicating the determination of the flow speed but revealing novel behavior in dense colloidal shear flow.}

\begin{figure*}
\includegraphics[width=1\textwidth]{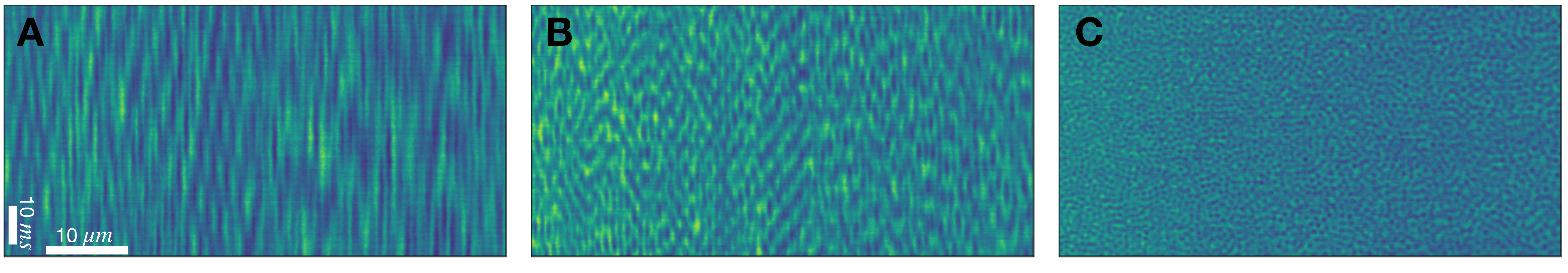}
\caption{ Representative space-time kymographs  generated from linescans (horizontal) stacked vertically taken in a sheared dense suspension of silica particles of diameter $1 \mu m$,  $\phi =0.52$, at different shear rates.  The spheres are not fluorescent, and a fluorescent dye is added to the solvent. A) $\dot\gamma = 10 s^{-1}$, $z=3 \mu m$,  B) $\dot\gamma = 20 s^{-1}$, $z=6 \mu m$,  C) $\dot\gamma = 30 s^{-1}$, $z=20 \mu m$.  Each image consists of 512 scans, for an elapsed time of 65 ms. }
\label{constant_SR}
\end{figure*}

\textbf{Validation:} To verify the accuracy of the linescan approach under these conditions, we performed tests on a moderately dense suspension, volume fraction $\phi =0.52$, where the viscosity is only weakly non-Newtonian, under constant shear rate conditions.  Figure \ref{constant_SR} shows three representative kymographs, each generated from 512 scans (65 ms total time), showing the range of speeds from the slowest ($\dot\gamma = 10 s^{-1}$; $z=3 \mu m$) to the fastest ($\dot\gamma = 30 s^{-1}$; $z=20 \mu m$). The wide range of transit times is evident, as is the appearance of regions of hexagonal ordering, which will be considered in the next section. 
Rather than identifying individual particles, we can quantify the length and time scales of the intensity variations using correlation analyses, with faster  flow producing more rapid fluctuations along the time axis.  In particular, as detailed in Methods, we calculate the speed of the flow by the rate of decay of the autocorrelation in the time direction (vertical in Fig. \ref{constant_SR}). This produces a very precise measurement of the fluctuation timescale, but conversion to speed requires knowing the length scale associated with the decay in the density autocorrelation.  At the volume fractions considered here ($\phi \sim 0.5$) that is of order the particle diameter, but the exact value is sensitive to the local structure of the suspension.  Thus, as discussed in the Methods section, there is some uncertainty in the absolute value of the measured speeds, and likely some correlation between the local order and the calculated speed.  In future work we plan to more fully characterize these effects with further experimental tests and comparisons with computer simulations.  

\begin{figure*}
\includegraphics[width=1.\textwidth]{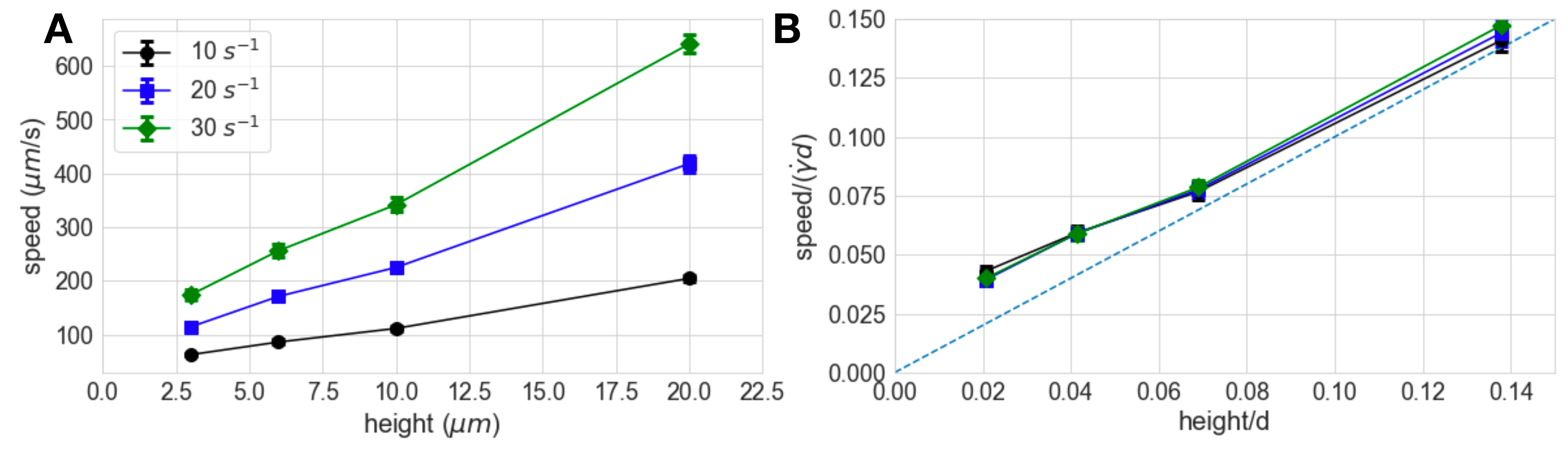}
\caption{A) Flow profiles for constant shear rate generated from linescan data at different heights   ($\phi =0.52$), at different constant shear rates. B) Particle speeds scaled by the speed of the top plate of the rheometer, $\dot\gamma d$, as a function of height above the bottom surface, scaled by the rheometer gap. Error bars are standard deviations (N=900) .  Dashed line corresponds to purely affine flow with no wall slip. }
\label{SR_profiles}
\end{figure*}

We acquired a series of 900 kymographs at heights above the bottom boundary of $z={3, 6, 10, 20 }\mu \rm{m}$, for constant applied shear rates of$\dot\gamma ={10, 20, 30} \rm{s}^{-1}$.  The resulting flow profiles are presented in Fig. \ref{SR_profiles}A, with error bars given by the standard deviation of each series. Figure \ref{SR_profiles}B shows the same data, with the speed scaled by the speed of the top plate, $v_p = \dot\gamma d$, where $d\approx 150 \mu m$ is the width of the rheometer gap at the imaging position.  (For the cone-plate geometry used here, the gap is a continuous function of radius, but changes relatively little over the $\approx 65 \mu m$ scan range.)   The data collapse, with no free parameters, to a unique, nearly linear profile that demonstrates the wide range of speeds over which the technique provides precise speed measurements.  
\NT{The deviation from an affine profile close to the wall likely arises from a combination of wall slip, layering, and ordering (because the ordering is strongest close to the wall and, as indicated above, spatial order will have an effect on the calculated speed).  Finally, we note that these measurements are taken at relatively high flow rates, so Brownian Motion or particle sedimentation are not  significant on the timescale of the decay of the intensity autocorrelation function (see Discussion).} 

\section{Results}
\begin{figure*}
\centering
\includegraphics[width=1.0\textwidth]{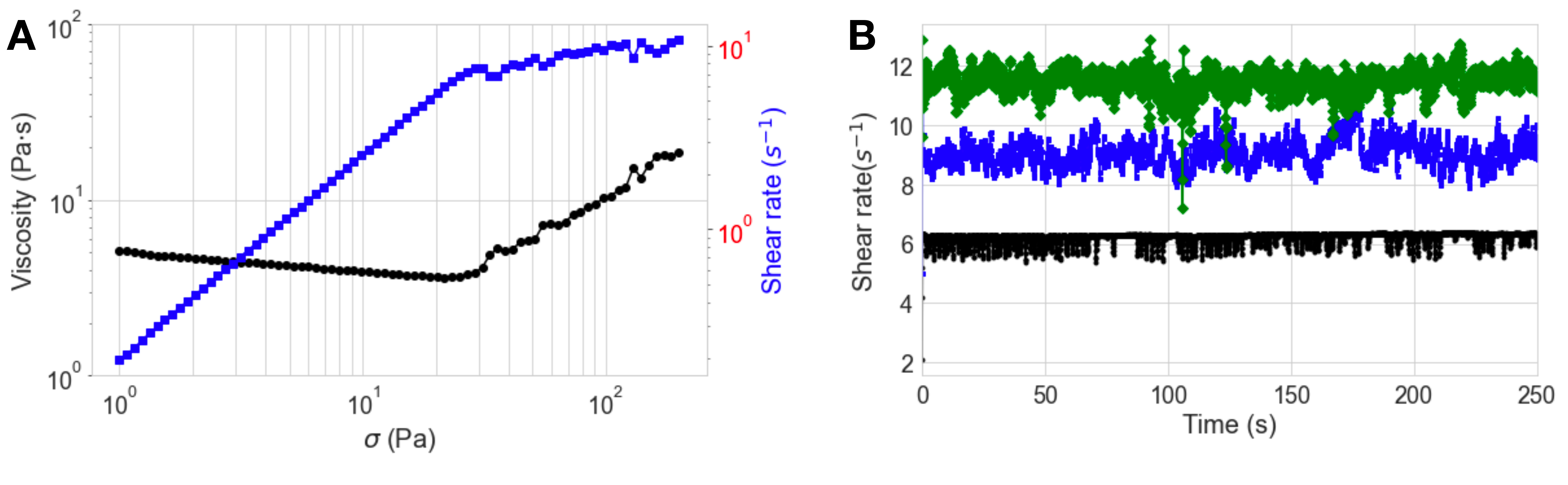}
\caption{A) Viscosity ({\color{black}$\CIRCLE$}) and shear rate ({\color{blue}$\blacksquare$ }) as a function of applied shear stress for the suspension used in this study, composed  of silica particles of diameter $1.0 \mu m$,  $\phi \approx 0.55$. B) Representative plots of shear rate vs. time for a constant applied stress of 20 (bottom), 50 (middle) and 100 (top) Pa.  Specifically, the time series shown correspond to the kymographs reported at $z=3 \mu m$. }
\label{flow_curve}
\end{figure*}

Here we report the results of a dense suspension of 1 $\mu m$ silica spheres at a volume fraction $\phi \approx 0.55$.  The  flow curve from bulk rheology for this sample is shown in Fig. \ref{flow_curve}, and exhibits behavior similar to previous reports of shear thickening in similar systems \cite{Morris:2020uy}, and in particular matches that of our previous results measuring local boundary stress fluctuations of the same system \cite{Rathee:2017ut}.  The viscosity shows substantial shear thinning (viscosity decreasing as applied shear stress is increased) until a critical stress of $\sigma _c \approx 20$ Pa, at which point strong shear thickening is observed. The transition occurs at a shear rate of $\dot\gamma _c \approx 10 s^{-1}$, which corresponds to a Peclet number $Pe= \dot\gamma    r^2  /D_o \approx 500$, where $D_o= k_BT/6 \pi \eta r$ is the diffusion coefficient for an isolated sphere of radius $r$ in a solvent of viscosity $\eta$
\cite{Vermant:2005tb, Morris:2009wo, Lettinga:2016wf} (approximately 80 $mPa\cdot s$ for the 80:20 glycerol:water mixture used here.  Note that the high concentrations used here substantially modify particle diffusion).
Figure \ref{flow_curve}B shows shear rate vs. time for constant applied stresses of 20, 50, and 100 Pa, corresponding to the conditions for the kymographs reported below. The increasing shear rate fluctuations with increasing stress is characteristic of a suspension showing strong shear thickening \cite{Morris:2020uy}.

\subsection{Order and density fluctuations before the onset of shear thickening} 
\begin{figure*}
\includegraphics[width=1\textwidth]{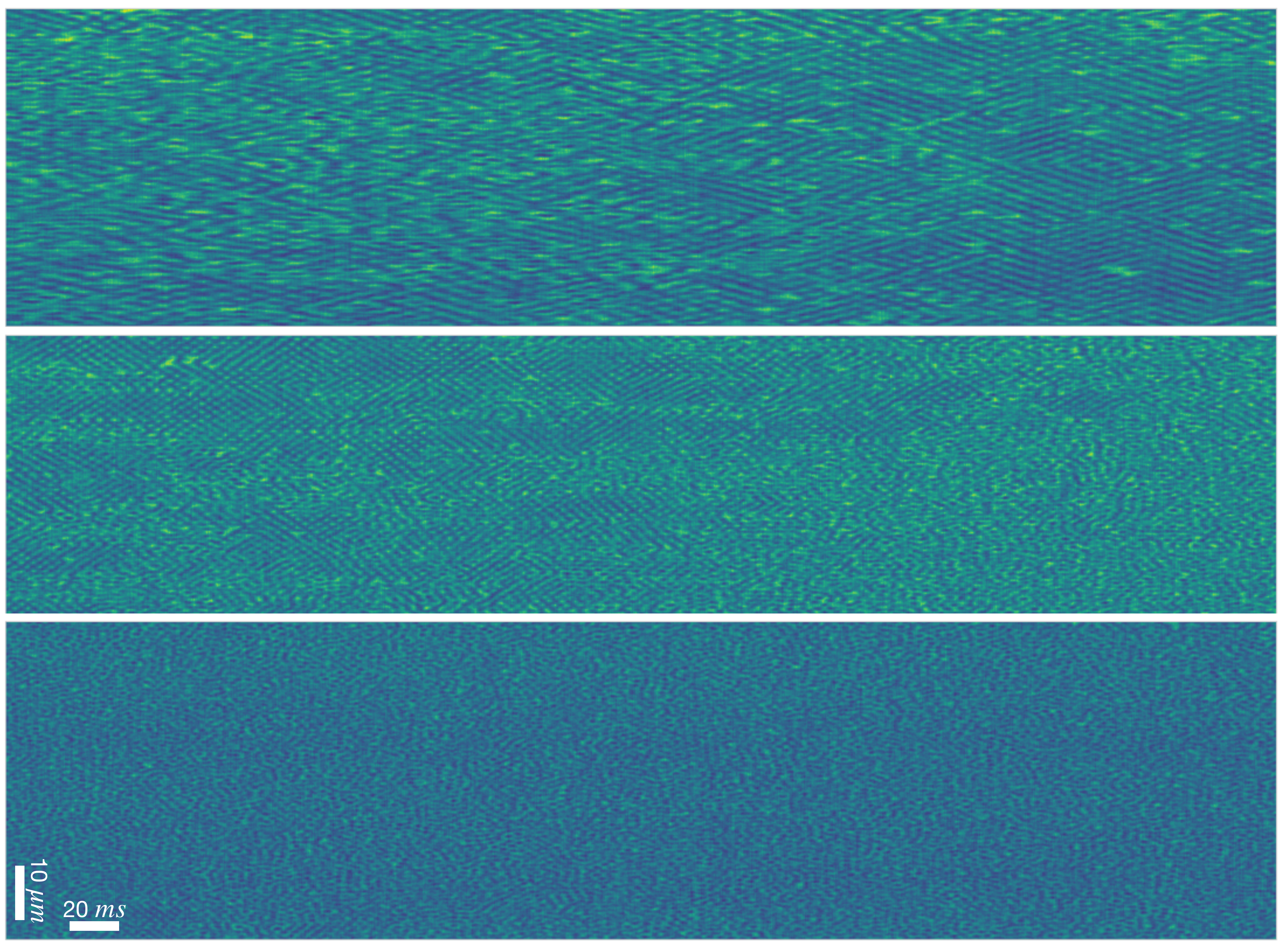}
\caption{  Representative space-time kymographs  generated from linescans (vertical) stacked horizontally at 20 Pa constant applied stress at heights of 3$\mu m$ (top), 10 $\mu m$ (middle), and 20 $\mu m$ above the bottom surface. Each image consists of 4096 scans, for an elapsed time of 520 ms. (Note that the orientation is different than Fig. \ref{constant_SR}, with time on the horizontal axis.) }
\label{20Pa-images}
\end{figure*}

As summarized in Section \ref{intro}, the shear thinning observed for applied stress $\sigma < \sigma _c$ is generally attributed to shear induced ordering, where organization of the monodisperse spheres into layers, and hexagonal ordering with the layers, reduces the viscosity by reducing collisions as the layers slide past one another.   Figure \ref{20Pa-images} shows representative kymographs at $\sigma$ = 20 Pa, at three different heights.   At $z=3 \mu m$ above the bottom surface, the kymograph (top) reveals a very regular hexagonal lattice, elongated along the time axis because the particle speed is relatively small.  Defects in the hexagonal order tend to be extended in the flow (time) directions, e.g. at the top right of the  panel. The left side of the kymograph shows a region of relatively low order, which is uncommon at this height, as quantified below.   At $z=10 \mu m$ (middle), regions of hexagonal order show up as small islands in a sea of disorder, with a tendency to be  be longer in the time direction than in the spatial (vorticity) direction.  Particle extent and separations in the time direction are proportional to the flow speed, so comparing dimensions requires rescaling by the flow speed 
{(Fig. S1),}
or, more simply, counting the number of particle diameters in the different directions.  The interpretation of the kymographs as direct images of 2D particle arrangements presumes that the rate of particle rearrangement is relatively slow compared to the rate at which structures traverse the scan region.  The persistence of the hexagonal order suggests that this is a reasonable approximation.  

In addition to revealing the existence and shape of small islands of order, the kymographs show that the ordered regions are darker than regions of disorder.  This can be seen clearly in the islands present at  $z=10 \mu m$, where the overall intensity is lower in the islands than around them (Fig. \ref{20Pa-images}, middle), but is also evident at $z=3 \mu m$ (top), where the small regions of disorder are brighter than their surroundings.  Because the fluorescent dye is in the solvent, brighter regions correspond to a {\em lower} concentration of colloidal particles.  Below we quantify the correlation between order and low intensity, and show that is substantial at all heights at 20 Pa applied stress.  This observation is consistent with simulations of dense suspensions of monodisperse particles, where a correlation between local ordering and high particle concentration was observed \cite{Goyal:2022wa}.

Quantifying the degree of hexagonal order in the kymographs is complicated by several factors. The images generated by individual particles depend on the height of the particle relative to imaging position. As discussed above, the particles tend to order in layers, and the patterns will be different when the center of the particle layer is at the linescan height compared to when the space between the layers aligns with the scan.  A further complication arises from the fact that the 1 $\mu m$ diameter of the particles is only a little larger than the resolution of the optical microscope, so particle images are not well separated.  Finally, the distortion of the image in the flow direction depends on the speed of the flow, which fluctuates in time.  Despite these complications we have found that we can extract a robust quantitative measure of the flow speed and order in each kymograph by calculating the two-dimensional intensity autocorrelation function, $g(\vec{r})$, for the entire  kymograph, rescaling distances in the time direction by the flow speed, and then calculating the degree of hexagonal order present at $|\vec{r}|=r_o$, where $r_o$ is the first peak in the angle averaged correlation function $g(r)$.  This measure, described in more detail in Section \ref{methods}, produces a complex scalar that is analogous to the quantity $C_6$ used in \cite{Morris:2009wo}. We use that name here, but note that because of the complications described above, even a fully ordered kymograph will produce a  $C_6$ with a magnitude that is considerably less than unity (in what follows, $C_6$ refers to the magnitude of the complex order parameter).
\begin{figure*}
\includegraphics[width=1\textwidth]{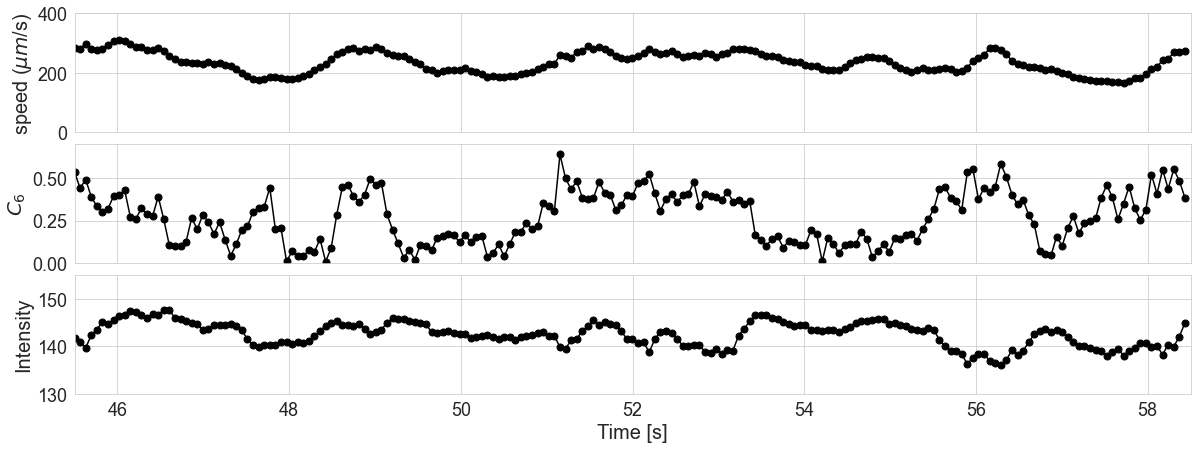}
\caption{  Section of time series of speed, hexagonal order, and intensity generated from kymographs at applied stress 20 Pa and height 10 $\mu m$. The first 8 data points correspond to the image shown in Fig. \ref{20Pa-images} (middle). }
\label{20Pa-timeseries}
\end{figure*}

Figure \ref{20Pa-timeseries} shows a portion of the timeseries of speed, $C_6$, and average intensity generated from kymographs at $z=10 \mu m$.  The first eight points in the time series correspond to the image shown in Fig. \ref{20Pa-images} (middle).  The decrease in order in the image (moving right to left) shows as a decrease in $C_6$ from $\approx 0.5$ to $< 0.3$.  Over the same region, the intensity increases modestly but significantly, as evident both in the image and the graph.  The full timeseries shows that there is a reasonably consistent anticorrelation between $C_6$ and intensity, consistent with the qualitative impression from small scale variations in individual images, but that there are also significant fluctuations that appear uncorrelated.    The correlations can be quantified by the Pearson correlation coefficient, and we find C($C_6$,intensity) = $\{ -0.19, -0.37, -0.34,-0.58\}$ at heights  $z= \{ 3, 6, 10, 20\} \mu m$.  The effects are particularly significant at the intermediate heights, where the fluctuations in $C_6$ are largest.  This negative correlation with intensity indicates that the ordered regions have a higher particle concentration, presumably as a consequence of more efficient packing.

A modest positive correlation can be observed  between order and speed, which is consistent with an ordering-induced increase in local velocity close to the wall, but is complicated by the fact that the correlation analysis used to determine the flow speed likely has a weak dependence on order (see Methods). Finally, we note that the orientation of the hexagonal order is such that intensity peaks and valleys always align in the flow direction, corresponding to a phase angle for the complex order parameter $\approx 0$ (data not shown). By contrast, the order seen at $\phi=0.52$, while considerably weaker (Fig. \ref{constant_SR}), has a phase angle $\approx \pi/6$ (Figure S3).

\begin{figure*}
\includegraphics[width=1\textwidth]{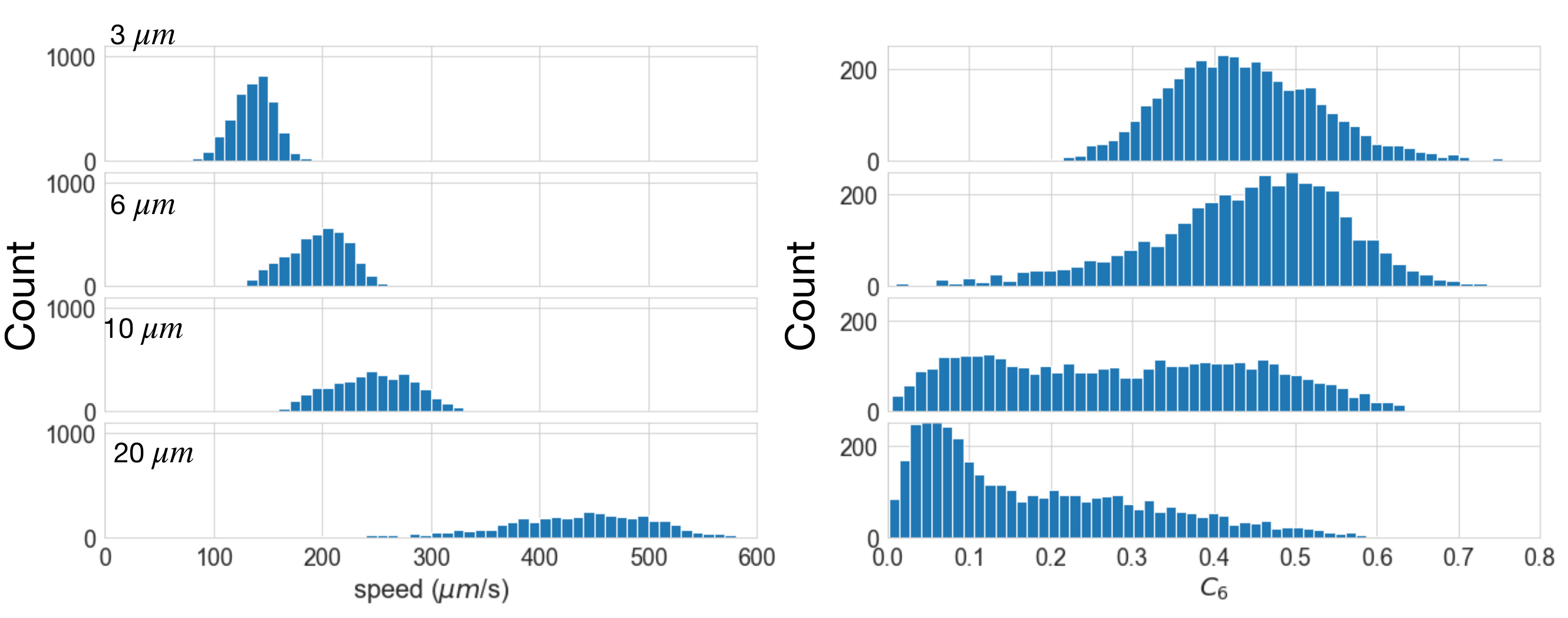}
\caption{Histograms of speed and order generated from kymographs at applied stress 20 Pa (Each height:  3900 kymographs, 65ms each).}
\label{20Pa-histograms}
\end{figure*}

\begin{figure*}
\includegraphics[width=1\textwidth]{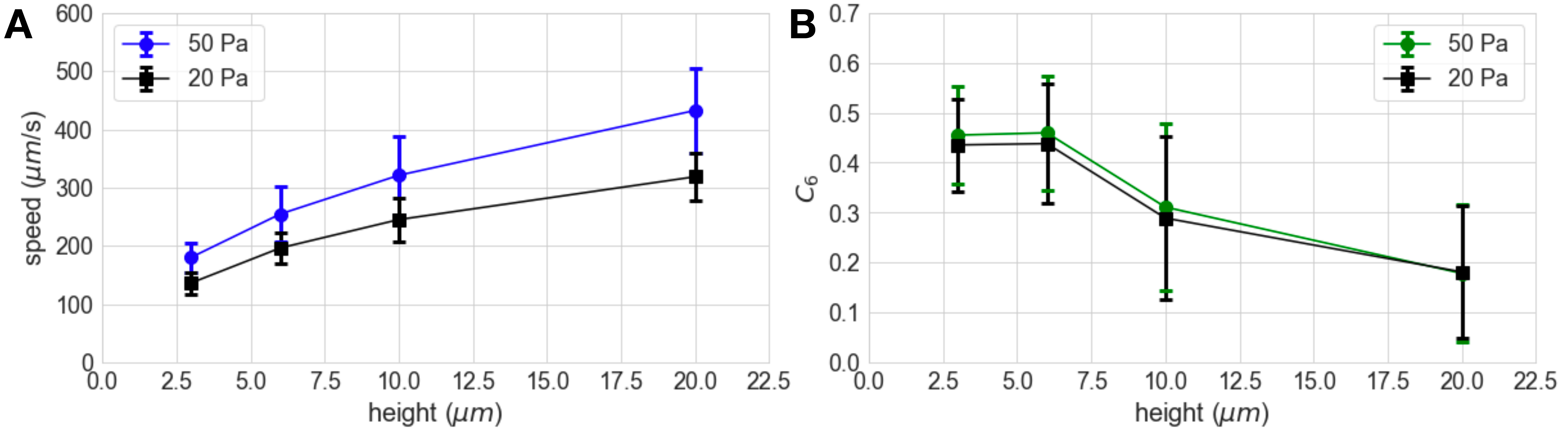}
\caption{Profiles of average velocity and order for 20 and 50 Pa applied stress. Error bars are standard deviations (N=3900).}
\label{20Pa-profiles}
\end{figure*}

The change in flow speed and order with distance from the bottom surface can be seen most clearly from histograms, as shown in Fig. \ref{20Pa-histograms}.  The speeds increase with height, as expected, with a significant increase in the spread of speeds that is roughly proportional to the speed increase.  Specifically, the ratio of the standard deviation to the mean speed is $\{0.14,0.19,0.21,0.17\}$ for heights $z=\{3,6,10,20\} \mu m$, respectively.  
The slightly higher fractional spread at 6 and 10 $\mu m$ likely arises from the coupling between order and speed mentioned above. The change in order with height is much more dramatic, showing uniformly high order $3 \mu m$ above the bottom surface, occasional disordered regions at $6 \mu m$, roughly equal order and disorder at  $10 \mu m$, and mostly disordered at $20 \mu m$.   
This behavior is summarized in Fig. \ref{20Pa-profiles}, which shows the average and standard deviations for speed and order as a function of height.  Figure  \ref{20Pa-profiles} also includes the data from 50 Pa applied stress, which shows very similar trends.  
The increase in curvature of the flow profile close to the boundary indicates a lower local viscosity.  Boundary-induced ordering and an associated viscosity decrease in dense suspensions has been seen previously, but the high speed imaging approach used here enables us to quantify spatio-temporal fluctuations in previously inaccessible regimes.

\subsection{Fluctuations associated with shear thickening} 
\begin{figure*}
\includegraphics[width=1.0\textwidth]{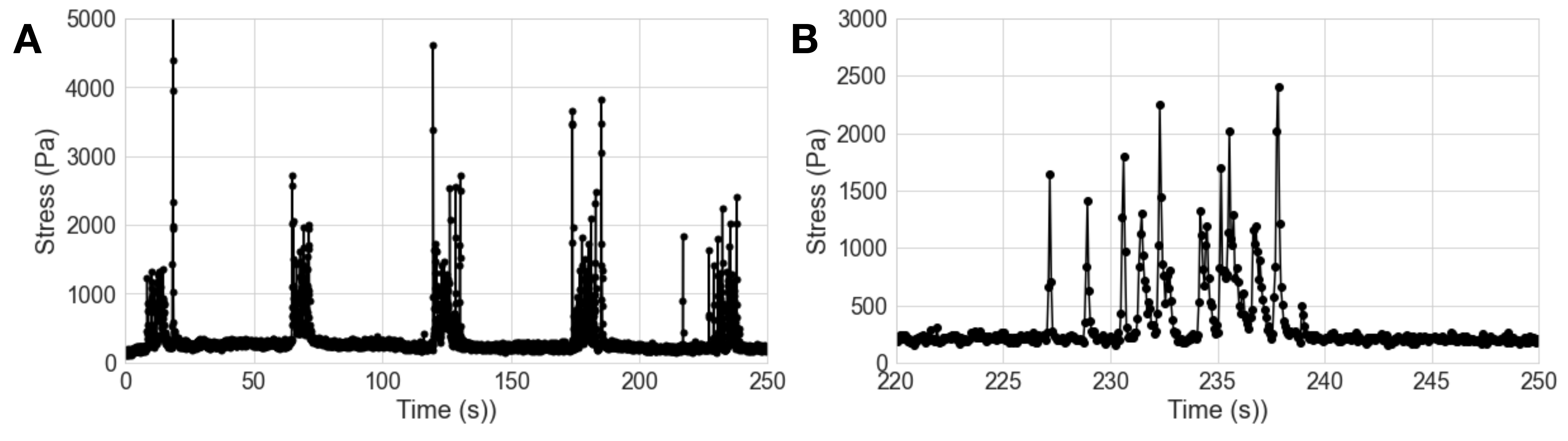}
\caption{A) Boundary stress (component in the flow direction) for 100 Pa applied stress. B) Expanded view of one cluster of events from data show in (A).}
\label{100Pa-BSM}
\end{figure*}

At higher applied stresses, the suspension shows substantial shear thickening and the nature of the fluctuations change dramatically.  We have previously shown that shear thickening is associated with large fluctuations in stresses at the boundary of the sheared suspension \cite{Rathee:2017ut,Rathee:2020un, Ganapathy:2020aa, Rathee:2020wi, Rathee:2022vy}, and specifically for the particles used here shear thickening is associated with a proliferation of the regions of high stress that propagate in the flow direction with approximately half the speed of the top plate ($v_p = \dot\gamma d$) \cite{Rathee:2017ut}.  Here local stress measurements at the bottom boundary are limited to a small region (62x62 $\mu m ^2$) but still show clustered spikes of high stress that propagate in the flow direction, plotted in figure \ref{100Pa-BSM}A.  The intermittent spikes are similar in magnitude and duration to those previously observed, but the clustering was not reported previously.  The connection with those and other results is discussed below, but for what follows the important observation is that the boundary stress shows regularly spaced clusters of intermittent high stress spikes. Figure \ref{100Pa-BSM}B shows an expanded view or one cluster of events, clearly showing the presence of large discrete spikes in the boundary stress with durations $\approx 100-500$ ms (consistent with all time series reported, the data points are separated by 65 ms).

\begin{figure*}
\includegraphics[width=1\textwidth]{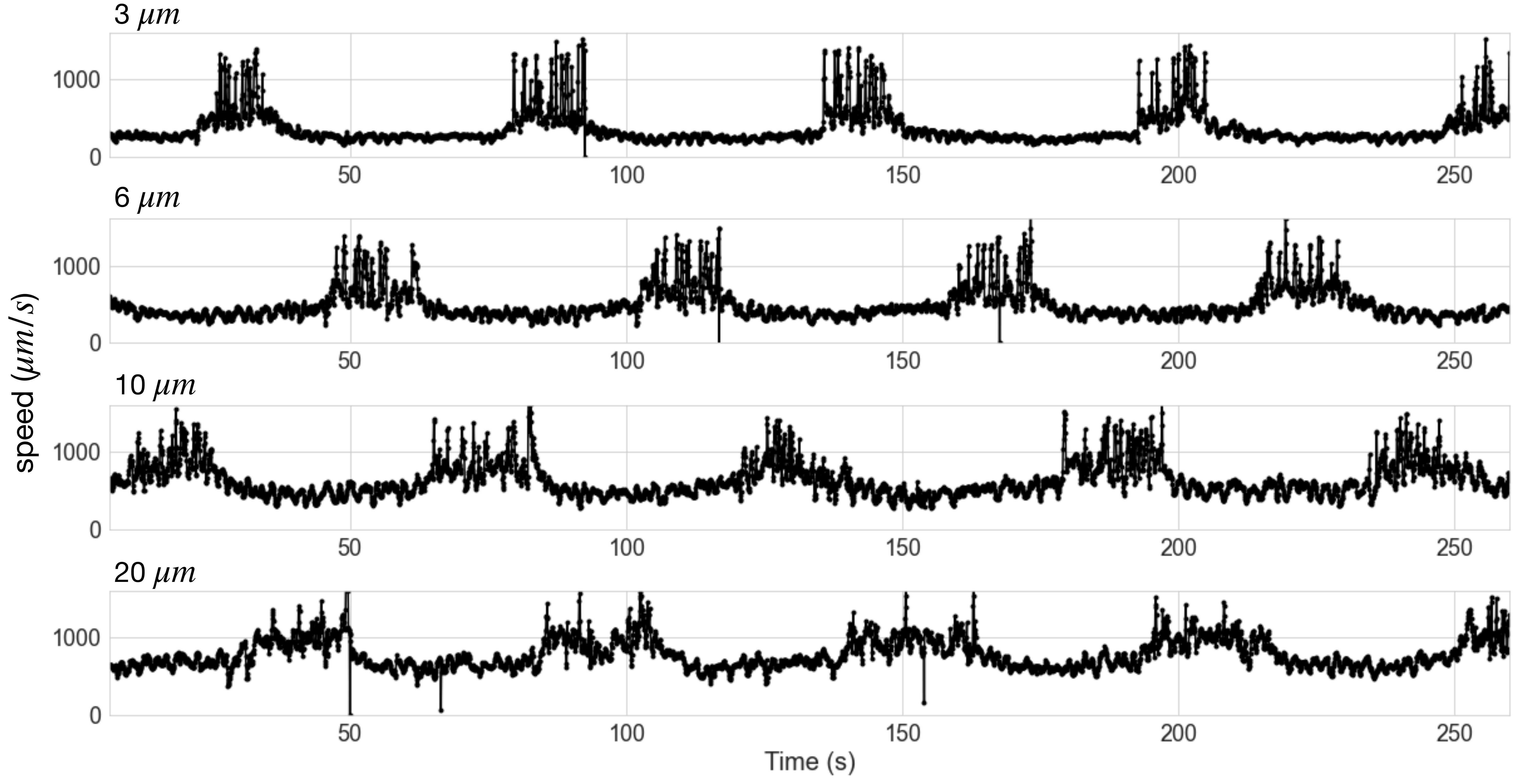}
\caption{Full timeseries of speeds for 100 Pa applied stress at heights of 3, 6, 10, and 20 $\mu m$.}
\label{100Pa-velocity}
\end{figure*}

Figure \ref{100Pa-velocity} shows the velocity time series for 4 different heights at 100 Pa applied stress.  The clustering of intermittent events is consistent in every data set. (Each time series is a different measurement run, and is different from the run that produced the boundary stress shown in Fig. \ref{100Pa-BSM}.) Thus we can be confident that the intermittent spikes in speed represent the flow fluctuations that are associated with the intermittent high boundary stress spikes.  Interestingly, the spacing between the clusters, $\sim 65 s$, is roughly 1/2 of the rotation period for the rheometer tool, suggesting that the clusters reflect stable features propagating in the flow direction with speed $v_p/2$. 

\begin{figure*}
\includegraphics[width=1\textwidth]{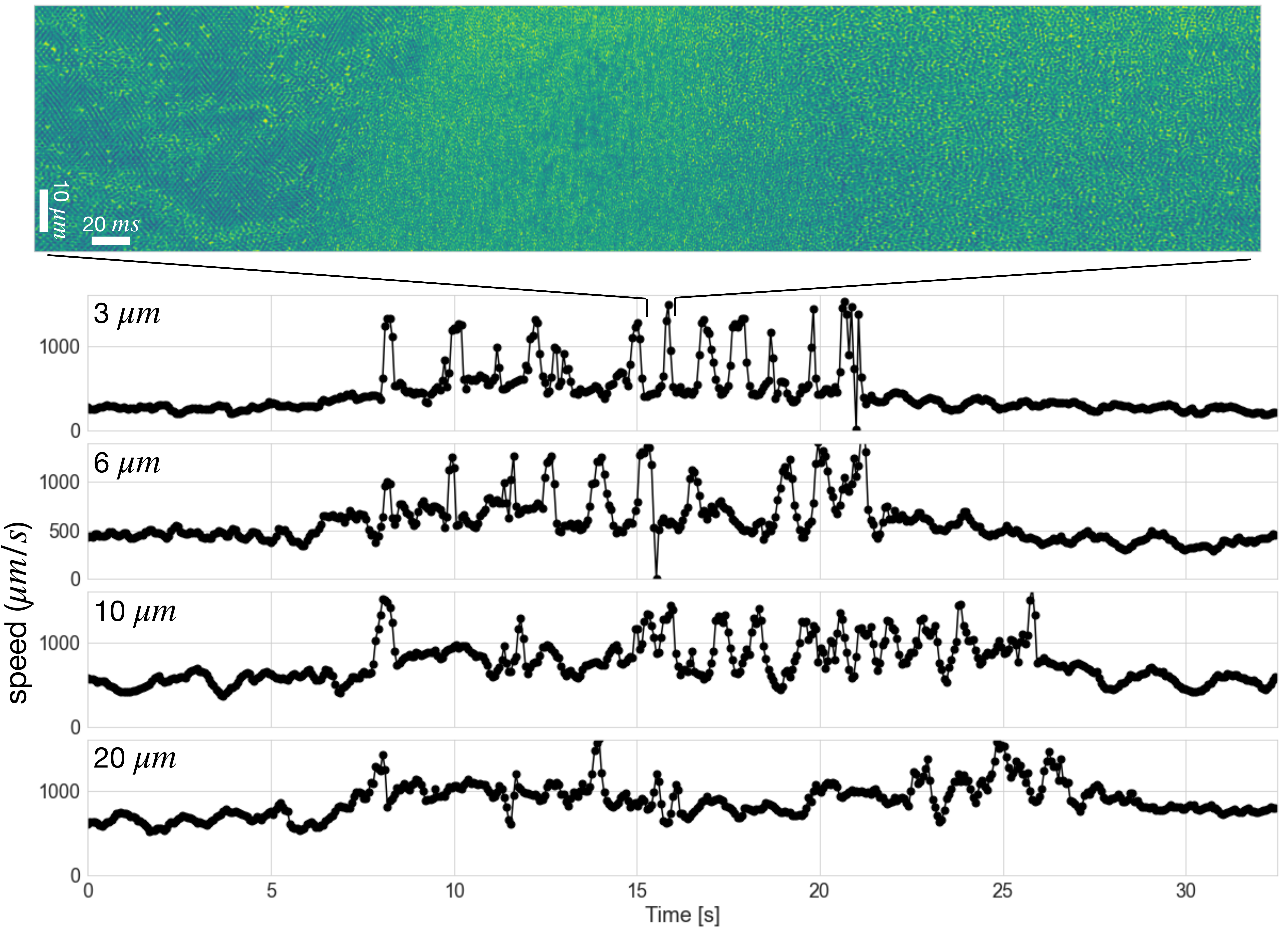}
\caption{Timeseries of speed for 100 Pa applied stress at heights of 3, 6, 10, and 20 $\mu m$ during one burst of high speed events.  The image shows a kymograph from a single event. The initial times are shifted so the bursts start at approximately the same time.}
\label{100Pa-events}
\end{figure*}

Figure \ref{100Pa-events} shows the time series for a cluster of spikes at each observed height, with the time axis shifted so that in each set the spikes start at the same time.  The kymograph at the top of the figure, showing  a single speed spike at $z=3 \mu m$, indicates that before the spike, the particles exhibit a high degree of hexagonal ordering, consistent with the behavior observed at this height at lower applied stress. The transition to high speed flow is evidenced by a remarkably sharp boundary that extends primarily along the vorticity direction, but with some meandering in the flow direction, consistent with the fluctuations in boundary stress reported previously \cite{Rathee:2017ut}. A complete absence of ordering and a significant increase in average intensity marks the period of high speed flow, suggesting a region of reduced particle concentration, although there are correlations between speed and intensity introduced by the instrumentation, and further calibration will be required to separate measurement effects from concentration when the speed variation is large.  The transition back to lower speed flow is also quite rapid, but the hexagonal ordering takes longer to recover.  These features are consistent across the high speed events, as shown by the time series of speed, $c_6$, and intensity at $3 \mu m$ (Figure S2).

Also evident in Fig. \ref{100Pa-velocity} is that the speed differential between the spikes and the background is smaller at larger $z$, indicating that the events represent a more dramatic speed-up close to the bottom boundary.  Finally, the clusters appear to be more spread out at larger $z$.  Since the periodicity is independent of $z$, the propagation speed of the clusters should be independent of $z$, so the spreading of the clusters with height suggests that the spatial extent of the region of high stress increases with depth.  

\begin{figure*}
\includegraphics[width=1\textwidth]{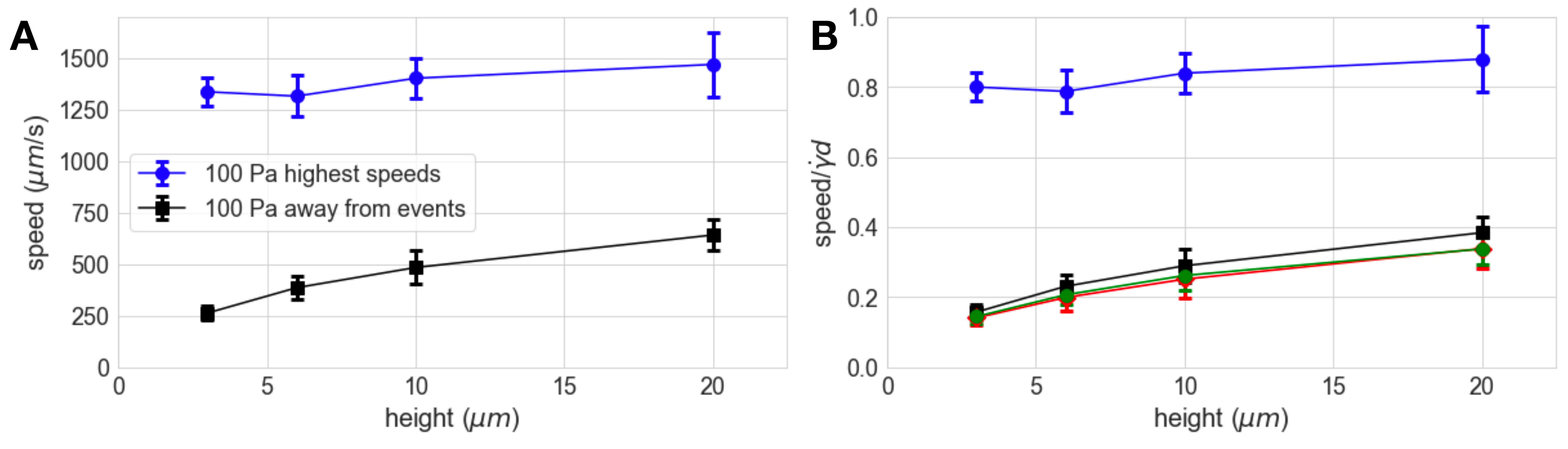}
\caption{A) Average speed at different heights from portions of the time series away from the high speed events, and average speed for the highest speeds in each time series (N=60).  B)  Data in (A) scaled by the speed of the top plate, with the scaled profiles from 20 and 50 Pa for comparison (Fig. \ref{20Pa-profiles}A) .}
\label{100Pa-profiles}
\end{figure*}

We can use the clustering of high speed events in the velocity time series to quantify the region of ``normal'' flow, away from the events (the specific segmentation is shown in Figure S3).  Figure \ref{100Pa-profiles}A shows the flow profile obtained during those time periods, and the overall shape closely matches that seen at lower stresses (Fig. \ref{20Pa-profiles}).  The agreement is made more evident by scaling the measured speeds by the average shear rate as reported by the rheometer (Fig. \ref{100Pa-profiles}B), where all three profiles overlap, with no free parameters. Something very different happens during the high speed events.  We find empirically that selecting the highest 60 speeds from the 3900 measurement points provides a reasonable measure of the peak speeds during the high speed events (Figure S4), and we find that the average of those top speeds is nearly height independent (Fig. \ref{100Pa-profiles}). Scaling the peak speeds by average the speed of the top plate shows that the height-independent speed is on the order of $v_p$, consistent with a solid jammed aggregate moving with the speed of the top plate, perhaps with some slip.  
\NT{(Note however, as discussed below, that the high speed events are quite likely associated with substantial non-affine flows, and thus it is possible that flows in the gradient direction contribute to the decay of the correlation and thus to the calculated speed.) }
Imaging of the bottom layer of particles confirms that the the high speed fluctuations are associated with large slip of that layer.  
Those apparent aggregates are intertwined with regions of the suspension where the particles are flowing much slower (Fig. \ref{100Pa-events}), which is inconsistent with a single solid aggregate.  These results provide the first direct measurements of the structure of the velocity field inside of high stress fluctuations in shear thickening suspensions.

\section{Discussion}

\textbf{Order and concentration fluctuations before the onset of shear thickening:} The dynamics revealed by high speed linescan imaging presented above are consistent with previous observations of boundary induced ordering in sheared dense monodisperse suspensions, but provide a more detailed picture of the spatiotemporal dynamics and demonstrate the presence of a strong connection between ordering and concentration fluctuations.  Planar boundaries enhance layering in colloidal suspensions, and the layering can reduces viscous dissipation, producing shear banding \cite{Mackay:2014ua,Kucuksonmez:2020vl, Goyal:2022wa}.  Hexagonal ordering of monodisperse particles within layers of sheared suspension has also been seen in a variety of circumstances, starting with the seminal work of Ackerson and coworkers. \cite{Ackerson:1990uh,Chen:1992ur}. 
Particularly relevant to this work, an imaging study by Wu et al. observed fluctuation in hexagonal ordering during the process of shear-induced melting in a confined suspension arising from the nucleation of localized domains that temporarily lost and regained their ordered structure \cite{Wu:2009vi}.  Similar behavior has been reproduced in computer simulations \cite{Mackay:2014ua,Myung:2013uf}. Recent simulation results reported that defects in crystalline order are associated with local decrease in particle concentration \cite{Goyal:2022wa}. 
It is perhaps not surprising that the interplay between shear flow, crystal nucleation, and shear-induced crystal breakup produces a complex phase diagram with complex dynamics \cite{Lettinga:2016wf}.  Here we show that for a monodisperse suspension at high packing fraction and high Peclet number ($Pe \sim 500$), the transition from mostly ordered to mostly disordered occurs $\sim 10$ particle diameters from the boundary (Fig. \ref{20Pa-profiles}), and is associated with rapid temporal fluctuations in order, speed, and concentration (Fig. \ref{20Pa-timeseries}).  The concentration fluctuations are particularly significant, because the local increase in particle concentration seen in ordered domains requires an compensating decrease in solvent concentration, which implies the existence of local fluid migration.  These results indicate that an accurate quantitative model for ordering fluctuations in sheared monodisperse suspensions at high Peclet number must include the flow and pore pressure fields responsible for relative flow between the particulate and fluid phases
\cite{devaraj:2916aa,ONeill:2019tv,Maharjan:2021ud, Rathee:2022vy}.

\textbf{Fluctuations associated with high local stresses during shear thickening:}  The high speed imaging approach employed here has revealed that localized high boundary stresses are accompanied by large rapid speed increases  near the boundary, as well as a loss of order (Fig. \ref{100Pa-events}).  Recent simulation results showed that the transition from hydrodynamic to frictional interparticle interactions that is believed to underlie strong shear thickening \cite{Morris:2020uy} is associated with a disruption of layering and ordering \cite{Goyal:2022wa}.  It seems likely that we are observing a similar transition, from relatively low stress, lubricated particle interactions producing layered, ordered low viscosity flow, to high stress, frictional particle interactions producing high viscosity disordered flow in the regions that produce high boundary stress.  This transition appears to be remarkably sharp, with a boundary that is only a few particles wide (Fig. \ref{100Pa-events}, top).  Also remarkable is the observation that the particle speeds during these events are independent of depth (Fig. \ref{100Pa-profiles}).  This suggests that the particles are moving together, as a jammed aggregate, similar to the model proposed to explain propagating high normal stresses observed in cornstarch suspensions \cite{Gauthier:2021vg}.  
\NT{It is important to note, however, that the linescan measurements performed here do not distinguish between flow components perpendicular to the scan direction (i.e. in the flow or gradient directions). As discussed above, under usual flow conditions the component of the velocity in the flow direction, of order $\dot\gamma z$,  will  be large compared to any non-affine flows, but the transient jamming observed during shear thickening is clearly unusual, so more complicated large-scale non-affine flows are likely.  Further measurements, for example addinge measurements with linescans oriented in the flow direction, are needed to get a clearer picture of the flow profile during the high stress events.}

More generally, measurements of nearly affine flow away from the localized high stresses, both in this system and in cornstarch \cite{Rathee:2022vy}, combined with the rapid fluctuation in velocity measured during the clusters of events (Fig. \ref{100Pa-events}), suggests a complicated flow field.  Furthermore, although instrumental effect preclude a direct measurement of particle concentration during the high speed events, the longstanding connection between frictional interactions and dilatancy \cite{Morris:2020uy}, evidenced by our observation of fluid migration associated with high stress fluctuations in cornstarch suspensions \cite{Rathee:2022vy}, suggests that relative flow between the particulate and fluid phases likely plays an important role in the observed speed fluctuations. 

These observations contribute to a growing body of evidence indicating that the shear thickening transitions typically involved complex spatiotemporal dynamics with structures propagating in the flow direction, including dilatant fronts \cite{Nakanishi:2012ux,Nagahiro:2016wu},  local deformations of the air-sample
interface at the edge of the rheometer tool \cite{Hermes:2016wj, Maharjan:2021ud}, local normal stresses in sheared cornstarch \cite{Gauthier:2021vg},  concentration fluctuations  appearing as periodic waves moving in the direction of flow \cite{Ovarlez:2020ul}, and high shear stress at the suspension boundary  \cite{Rathee:2017ut, Rathee:2020wi, Rathee:2020un, Rathee:2022vy}. The specifics of the dynamics vary considerably, presumably indicating a sensitivity to the details of the suspension and the measurement geometry.  This sensitivity is perhaps not surprising, given that shear thickening involves instabilities that can produce discontinuities in material parameters. The location of those discontinuities in a uniform extended system represent a broken symmetry, and thus in any physical realization will be very sensitive to the parameter variations (e.g. of shear rate, rheometer gap, distance from suspension boundary) that are present in all experimental systems. 

Here we have presented an initial application of a powerful new approach that reveals spatiotemporal dynamics of sheared dense suspensions with high spatial and temporal resolution.  With further testing and validation the approach has the potential to provide accurate and precise measurements of local speed, structure, and particle concentration, and should provide a new avenue to answer open questions about dense suspensions under flow.

\section{Methods}
\label{methods}

\begin{figure*}
\includegraphics[width=1\textwidth]{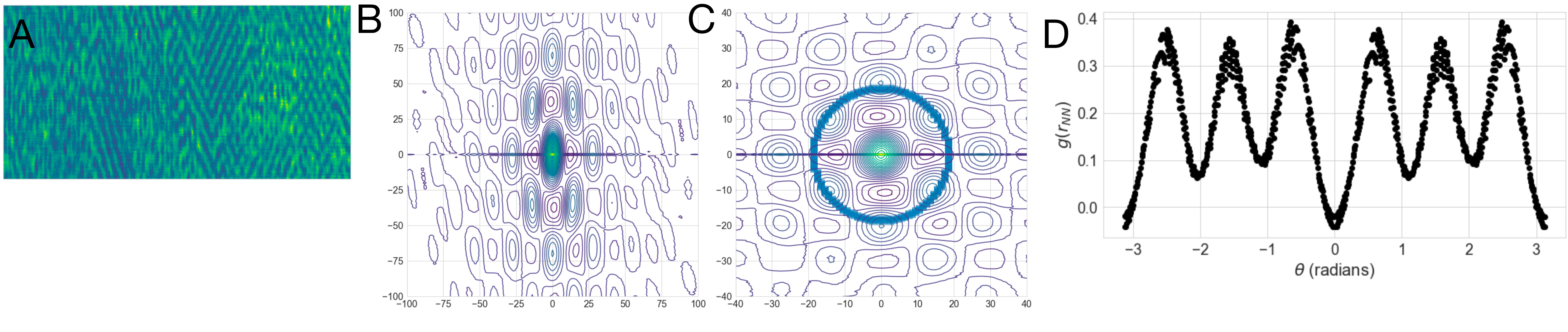}
\caption{  A) Example kymograph showing showing moderate hexagonal ordering.  B) Two-dimensional autocorrelation of (A), with the central region used to determine the peak widths highlighted. C) 2D autocorrelation with vertical distances rescaled so that the central peak is circularly symmetric.  The ring represents points whose distance from the origin is equal to the expected average inter-particle spacing.  D) Intensity as a function of angle for the ring of points identified in (C), resulting in a $C_6$ with magnitude 0.36 and angle 0.03 radians.}
\label{G(r)}
\end{figure*}

All experimental suspensions were composed of $0.9 \mu m$ silica beads (Bang's Lab) in an index-matched (80/20 v/v) glycerol/water mixture. For imaging purposes, fluorescein sodium salt was added to the suspension so that the unlabeled spheres could be imaged as dark spots in a fluorescent background. Rheological measurements were
performed on a stress-controlled rheometer (Anton Paar MCR 301)
mounted on an inverted confocal (Leica SP5) microscope \cite{Dutta:2013aa}
using a cone-plate geometry with a diameter 25 mm.  The linescan data was acquired with a 60X objective at a radius of 2/3 of the plate diameter, where the rheometer gap is $d \approx 145 \mu m$.\\
\textbf{Linescan analysis:} 
A typical data set is composed of a 1-2 million scans, with the scan direction aligned along the vorticity axis (perpendicular to the flow and the gradient).  For visualization, we typically divide the set into a series of 2D arrays, with one dimension (horizontal) given by the number of pixels per scan (here  1024), and the other dimension (vertical) given by the chosen number of scans per image, in this case 512 scans.  Fig. \ref{G(r)}A shows an example image generated from  $1 \mu$m diameter non-fluorescent spheres in a fluorescent background, 10 $\mu m$ above the bottom of the sheared suspension, similar to those shown in Fig.\ref{20Pa-images}.   Individual particles show up as dark ovals.  In principle, the vertical axis of each oval could be used to measure the speed of each particle.  In practice, in many places identifying individual particles is challenging, and we have found correlation analyses more reliable.  (This arises in part because we have no control of the position of particles relative to the plane of focus, and the image will include contributions from particles that are $\approx 1 \mu$m above or below the focal plane).  Note also that the orientation of the ovals provides a measure for the component of the flow in the vorticity direction.  Here that component is always small, and is not included in our analysis. An alternative approach, not employed here, would be to orient the scan line along the flow direction.  This would enable very precise measurements of flow speed, as particles move along the line, but would not reveal spatial structure in the flow-vorticity plane.

\begin{figure*}
\includegraphics[width=1\textwidth]{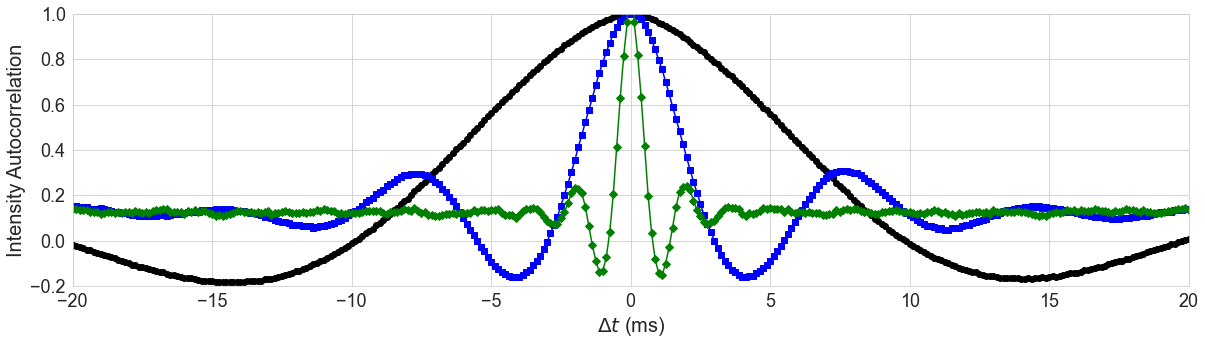}
\caption{
Normalized intensity autocorrelation function $g(\Delta t)$ as defined in Methods for the three kymographs shown in Fig. 2,  $\dot\gamma = 10 s^{-1}$ , $z=3 \mu m$ ({\color{black}$\CIRCLE$}),  $\dot\gamma = 20 s^{-1}$ , $z=6 \mu m$ ({\color{blue}$\blacksquare$ }),  and  $\dot\gamma = 30 s^{-1}$, $z=20 \mu m$  ({\color{green}$\blacklozenge$}). The width of the central peak and the position of the first minimum are proportional to the flow speed, for fixed positional configurations, but  both measures are have a some sensitivity to the presence of positional order.}
\label{g_dt}
\end{figure*}

One approach to extract a characteristic transit time from the linescan intensity data, $I(x,t)$, is to calculate the autocorrelation in the time direction, $g(\Delta t) = <\delta I(x,t)\delta I(x,t+\Delta t)>_{x,t}/<\delta I (x,t) ^2>_{x,t}$, where $\delta I (x,t)=I(x,t)-<I(x,t)>_{x,t}$.  The range of $x$ and $t$ to include in the averages can be varied depending on the spatial and temporal resolution required by the effective particle size. Slower shear rates require averaging over larger time windows so as to capture at least one full particle transit. For the conditions in this study we found that 512 scans provides enough data for robust correlation analysis while still allowing adequate temporal resolution. 
Figure \ref{g_dt} shows $g(\Delta t)$ for the three images displayed in Fig. \ref{constant_SR}. 

The location of the first minimum in the autocorrelation, $t_{min}$ can be precisely identified algorithmically, and provides a measure that is proportional to the speed of the flow.  
We expect that $v_{flow}=l_{flow}/2t_{min}$, where $l_{flow}$ is approximately equal to the spacing between particles in the flow direction.
More precisely, $l_{flow}$ should be equal to the first minimum in the average instatntaneous density autocorrelation calculated along the flow axis.  Unfortunately that quantity is unknown.  Assuming it does not change with time, $v_{flow} \propto 1/t_{min}$.  Similarly, 
the minimum of the autocorrelation in the space direction provides a direct measure of the spacing in the $x$ (vorticity) direction, specifically $l_{vorticity}=2x_{min}.$  We find that this measure does not depend directly on the flow speed, but is quite sensitive to local hexagonal ordering.  In part because of this sensitivity, and in part because we are interested in quantitative measures of hexagonal order, we have instead employed a slightly more complex analysis approach based on the 2D autorcorrelation function, but we have found that the speed variations are essentially indistinguishable from those calculated with the 1D correlation analysis.

Using a 2D correlation analysis, we can quantify the hexagonal order that is evident in  Fig. \ref{G(r)}A.  It is important to remember that the image is not a snapshot of the two-dimensional particle arrangement, as the top of the image is generated at an earlier time than the bottom.  However, if the structure of the suspension evolves slowly compared to the time to generate an image, it will in fact be an accurate representation of the spatial arrangement of the particles, with the vertical separations expanded (or compressed) by a factor $\alpha=(s\cdot d)/v_{flow}$, where $s$ is the scan rate, $d$ is the actual spatial separation, and $v$ is the speed of the flow.  Figure \ref{G(r)}B shows a contour plot of the two-dimensional spatial autocorrelation, $G(\vec{r})$, of the image shown in A.  The sixfold symmetry is evident, as is the expansion in the vertical direction.  Note  that $G(\vec{r})$ is large along the horizontal axis 
\textcolor{red}{($y=0)$}.  
This is likely a consequence of fluctuations in the laser intensity on a sub-millisecond timescale. These fluctuations can be filtered out, but for our purposes it is sufficient to simply exclude the horizontal axis from our analysis of $G(\vec{r})$.  Fitting the central peak of $G(\vec{r})$ (highlighted region in Fig. \ref{G(r)}B) to a 2D Gaussian provides an accurate measure of the width in the flow direction, $\Delta y$ (number of scan lines)  which in inversely proportional to the speed, 
$v_{flow}=\delta \cdot s /\Delta y$.  
Determining the scale factor  $\delta$, the width of the central peak in $G(\vec{r})$ in physical units, is 
a significant
source of uncertainty in the measurement.  The particle diameter is known precisely, but $\delta$ also depends on the particle separation, and is also somewhat sensitive to the degree of order.  We have estimated $\delta$ from looking at $G(x)$ (the correlation along the scan direction) in flows with no detectable order, where we find that the peak width is 0.2 times the distance to the first peak in $G(x)$. Assuming  the first peak in G(x) is at a location $d_{peak}=d_{particle}/\phi^{-1/3}\approx 1.2 *d_{particle}$, we find $\delta \approx 1.2 d_{particle}/5$. This conversion enables us to provide speed measurements in physical units, but further tests will be required to validate the accuracy of the results.  For the measurements described here, the $\sim 10\%$ uncertainty in the exact value of the proportionality constant does not impact the conclusions, but  systematic effects of ordering on $\Delta y$ may impact some results, e.g. the apparent correlation between order and speed in the measurements at 20 Pa.  

Using the measured width $\Delta y$, we can rescale the flow direction to produce a symmetric $G(\vec{r})$. Figure \ref{G(r)}C shows a contour plot of the rescaled $G(\vec{r})$, and thus represents our measure of the average spatial structure of the imaged region.  In the results presented below, we are primarily interested in measuring the degree of hexagonal order in the suspension. We can quantify this by measuring the intensity as a function of angle at a distance $r_{NN}$ from the origin, where $r_{NN}$ is the position of the first maximum in $G(r)$, approximately the average nearest neighbor separation.   The points included in the calculation of are highlighted in Fig. \ref{G(r)}, and the resulting data $G(r_{NN},\theta)$ is shown in Figure \ref{G(r)}D. A complex scaler $C_6$ representing the hexagonal order can be calculated according to


\begin{equation}
 C_6 = \frac{\int _0^{2\pi} G(r_{NN}, \theta)\exp^{-i6\theta} d\theta}{\int _0^{2\pi} G(r_{NN}, \theta)d\theta}.
\end{equation}

\textbf{Boundary stress microscopy}. BSM measurements (Fig. \ref{100Pa-BSM} were performed as described in  \cite{Rathee:2017ut,Rathee:2020wi}, but employing elastic films of relatively high modulus ($G\sim 1 MPa$) to minimize possible effects of boundary compliance.  Briefly, elastic films  thickness 50 $\pm 3 \mu$m  were deposited by spin coating PDMS (Sylgard 184; Dow Corning) and a curing
agent  on 40 mm diameter glass cover slides (Fisher Sci) that were cleaned thoroughly by plasma cleaning and rinsing
with ethanol and deionized water \cite{Rathee:2017ut}.    After deposition of PDMS, the slides
were cured at 85 $^{o}$C for 2 hours. After curing, the PDMS was
functionalized with 3-aminopropyl triethoxysilane (Fisher Sci) using
vapor deposition for $~$40 min. Carboxylate-modified
fluorescent spherical beads of radius 0.5 $\mu$m with
excitation/emission at 520/560 nm were attached to the PDMS
surface. Before attaching the beads to functionalized PDMS, the beads
were suspended in a solution containing PBS solution
(Thermo-Fisher). The concentration of beads used was 0.006 $\%$
solids. 
The surface stresses at the interface were calculated using an extended traction force technique and codes given in ref. \cite{Style:2014aa}.  Taking the component of the surface stress in the flow (velocity) direction, we obtain the  scalar field $\sigma_{BSM}(\vec{r},t)$, representing the spatiotemporally varying surface stress. The data shown in Fig. \ref{100Pa-BSM} are a spatial average of each field, $<\sigma_{BSM}(\vec{r},t)>_t$.

\section{Acknowledgements}
The authors thank  Peter Olmsted and Emanuela Del Gado for helpful discussions. This work was supported by the 
National Science Foundation (NSF) under Grant No. DMR-1809890. J. S. U. is supported, in part, by the Georgetown 
Interdisciplinary Chair in Science Fund.


\bibliography{manuscript.bib}

\begin{thebibliography}{49}
\expandafter\ifx\csname natexlab\endcsname\relax\def\natexlab#1{#1}\fi
\expandafter\ifx\csname bibnamefont\endcsname\relax
  \def\bibnamefont#1{#1}\fi
\expandafter\ifx\csname bibfnamefont\endcsname\relax
  \def\bibfnamefont#1{#1}\fi
\expandafter\ifx\csname citenamefont\endcsname\relax
  \def\citenamefont#1{#1}\fi
\expandafter\ifx\csname url\endcsname\relax
  \def\url#1{\texttt{#1}}\fi
\expandafter\ifx\csname urlprefix\endcsname\relax\def\urlprefix{URL }\fi
\providecommand{\bibinfo}[2]{#2}
\providecommand{\eprint}[2][]{\url{#2}}

\bibitem[{\citenamefont{Vermant and Solomon}(2005)}]{Vermant:2005tb}
\bibinfo{author}{\bibfnamefont{J.}~\bibnamefont{Vermant}} \bibnamefont{and}
  \bibinfo{author}{\bibfnamefont{M.~J.} \bibnamefont{Solomon}},
  \bibinfo{journal}{Journal of Physics: Condensed Matter}
  \textbf{\bibinfo{volume}{17}}, \bibinfo{pages}{R187 R216}
  (\bibinfo{year}{2005}), ISSN \bibinfo{issn}{0953-8984}.

\bibitem[{\citenamefont{Morris}(2009)}]{Morris:2009wo}
\bibinfo{author}{\bibfnamefont{J.~F.} \bibnamefont{Morris}},
  \bibinfo{journal}{Rheologica Acta} \textbf{\bibinfo{volume}{48}},
  \bibinfo{pages}{909} (\bibinfo{year}{2009}), ISSN \bibinfo{issn}{0035-4511}.

\bibitem[{\citenamefont{Lettinga}(2016)}]{Lettinga:2016wf}
\bibinfo{author}{\bibfnamefont{M.~P.} \bibnamefont{Lettinga}}, pp.
  \bibinfo{pages}{81--110} (\bibinfo{year}{2016}).

\bibitem[{\citenamefont{Chen et~al.}(1992)\citenamefont{Chen, Zukoski,
  Ackerson, Hanley, Straty, Barker, and Glinka}}]{Chen:1992ur}
\bibinfo{author}{\bibfnamefont{L.~B.} \bibnamefont{Chen}},
  \bibinfo{author}{\bibfnamefont{C.~F.} \bibnamefont{Zukoski}},
  \bibinfo{author}{\bibfnamefont{B.~J.} \bibnamefont{Ackerson}},
  \bibinfo{author}{\bibfnamefont{H.~J.~M.} \bibnamefont{Hanley}},
  \bibinfo{author}{\bibfnamefont{G.~C.} \bibnamefont{Straty}},
  \bibinfo{author}{\bibfnamefont{J.}~\bibnamefont{Barker}}, \bibnamefont{and}
  \bibinfo{author}{\bibfnamefont{C.~J.} \bibnamefont{Glinka}},
  \bibinfo{journal}{Physical Review Letters} \textbf{\bibinfo{volume}{69}},
  \bibinfo{pages}{688} (\bibinfo{year}{1992}), ISSN \bibinfo{issn}{0031-9007}.

\bibitem[{\citenamefont{Holmqvist et~al.}(2005)\citenamefont{Holmqvist,
  Lettinga, Buitenhuis, and Dhont}}]{Holmqvist:2005aa}
\bibinfo{author}{\bibfnamefont{P.}~\bibnamefont{Holmqvist}},
  \bibinfo{author}{\bibfnamefont{M.~P.} \bibnamefont{Lettinga}},
  \bibinfo{author}{\bibfnamefont{J.}~\bibnamefont{Buitenhuis}},
  \bibnamefont{and} \bibinfo{author}{\bibfnamefont{J.~K.~G.}
  \bibnamefont{Dhont}}, \bibinfo{journal}{Langmuir}
  \textbf{\bibinfo{volume}{21}}, \bibinfo{pages}{10976} (\bibinfo{year}{2005}),
  ISSN \bibinfo{issn}{0743-7463}.

\bibitem[{\citenamefont{Wu et~al.}(2009)\citenamefont{Wu, Derks, Blaaderen, and
  Imhof}}]{Wu:2009vi}
\bibinfo{author}{\bibfnamefont{Y.~L.} \bibnamefont{Wu}},
  \bibinfo{author}{\bibfnamefont{D.}~\bibnamefont{Derks}},
  \bibinfo{author}{\bibfnamefont{A.~v.} \bibnamefont{Blaaderen}},
  \bibnamefont{and} \bibinfo{author}{\bibfnamefont{A.}~\bibnamefont{Imhof}},
  \bibinfo{journal}{Proceedings of the National Academy of Sciences of the
  United States of America} \textbf{\bibinfo{volume}{106}},
  \bibinfo{pages}{10564} (\bibinfo{year}{2009}), ISSN
  \bibinfo{issn}{0027-8424}.

\bibitem[{\citenamefont{Derks et~al.}(2009)\citenamefont{Derks, Wu, Blaaderen,
  and Imhof}}]{Derks:2009tc}
\bibinfo{author}{\bibfnamefont{D.}~\bibnamefont{Derks}},
  \bibinfo{author}{\bibfnamefont{Y.~L.} \bibnamefont{Wu}},
  \bibinfo{author}{\bibfnamefont{A.~v.} \bibnamefont{Blaaderen}},
  \bibnamefont{and} \bibinfo{author}{\bibfnamefont{A.}~\bibnamefont{Imhof}},
  \bibinfo{journal}{Soft Matter} \textbf{\bibinfo{volume}{5}},
  \bibinfo{pages}{1060} (\bibinfo{year}{2009}), ISSN \bibinfo{issn}{1744-683X}.

\bibitem[{\citenamefont{Richard and Speck}(2015)}]{Richard:2015tz}
\bibinfo{author}{\bibfnamefont{D.}~\bibnamefont{Richard}} \bibnamefont{and}
  \bibinfo{author}{\bibfnamefont{T.}~\bibnamefont{Speck}},
  \bibinfo{journal}{Scientific Reports} \textbf{\bibinfo{volume}{5}},
  \bibinfo{pages}{14610} (\bibinfo{year}{2015}), \eprint{1509.08687}.

\bibitem[{\citenamefont{Shereda et~al.}(2010)\citenamefont{Shereda, Larson, and
  Solomon}}]{Shereda:2010te}
\bibinfo{author}{\bibfnamefont{L.~T.} \bibnamefont{Shereda}},
  \bibinfo{author}{\bibfnamefont{R.~G.} \bibnamefont{Larson}},
  \bibnamefont{and} \bibinfo{author}{\bibfnamefont{M.~J.}
  \bibnamefont{Solomon}}, \bibinfo{journal}{Physical Review Letters}
  \textbf{\bibinfo{volume}{105}}, \bibinfo{pages}{228302}
  (\bibinfo{year}{2010}), ISSN \bibinfo{issn}{0031-9007}.

\bibitem[{\citenamefont{Cheng et~al.}(2012)\citenamefont{Cheng, Xu, Rice,
  Dinner, and Cohen}}]{Cheng:2012tt}
\bibinfo{author}{\bibfnamefont{X.}~\bibnamefont{Cheng}},
  \bibinfo{author}{\bibfnamefont{X.}~\bibnamefont{Xu}},
  \bibinfo{author}{\bibfnamefont{S.~A.} \bibnamefont{Rice}},
  \bibinfo{author}{\bibfnamefont{A.~R.} \bibnamefont{Dinner}},
  \bibnamefont{and} \bibinfo{author}{\bibfnamefont{I.}~\bibnamefont{Cohen}},
  \bibinfo{journal}{Proceedings of the National Academy of Sciences}
  \textbf{\bibinfo{volume}{109}}, \bibinfo{pages}{63} (\bibinfo{year}{2012}),
  ISSN \bibinfo{issn}{0027-8424}.

\bibitem[{\citenamefont{Xu et~al.}(2013)\citenamefont{Xu, Rice, and
  Dinner}}]{Xu:2013wt}
\bibinfo{author}{\bibfnamefont{X.}~\bibnamefont{Xu}},
  \bibinfo{author}{\bibfnamefont{S.~A.} \bibnamefont{Rice}}, \bibnamefont{and}
  \bibinfo{author}{\bibfnamefont{A.~R.} \bibnamefont{Dinner}},
  \bibinfo{journal}{The Journal of Physical Chemistry Letters}
  \textbf{\bibinfo{volume}{4}}, \bibinfo{pages}{3310} (\bibinfo{year}{2013}),
  ISSN \bibinfo{issn}{1948-7185}, \eprint{1310.1331}.

\bibitem[{\citenamefont{Mackay et~al.}(2014)\citenamefont{Mackay, Pastor,
  Karttunen, and Denniston}}]{Mackay:2014ua}
\bibinfo{author}{\bibfnamefont{F.~E.} \bibnamefont{Mackay}},
  \bibinfo{author}{\bibfnamefont{K.}~\bibnamefont{Pastor}},
  \bibinfo{author}{\bibfnamefont{M.}~\bibnamefont{Karttunen}},
  \bibnamefont{and}
  \bibinfo{author}{\bibfnamefont{C.}~\bibnamefont{Denniston}},
  \bibinfo{journal}{Soft Matter} \textbf{\bibinfo{volume}{10}},
  \bibinfo{pages}{8724} (\bibinfo{year}{2014}), ISSN \bibinfo{issn}{1744-683X}.

\bibitem[{\citenamefont{Villada-Balbuena
  et~al.}(2022)\citenamefont{Villada-Balbuena, Jung, Zuccolotto-Bernez,
  Franosch, and Egelhaaf}}]{villada:2022aa}
\bibinfo{author}{\bibfnamefont{A.}~\bibnamefont{Villada-Balbuena}},
  \bibinfo{author}{\bibfnamefont{G.}~\bibnamefont{Jung}},
  \bibinfo{author}{\bibfnamefont{A.~B.} \bibnamefont{Zuccolotto-Bernez}},
  \bibinfo{author}{\bibfnamefont{T.}~\bibnamefont{Franosch}}, \bibnamefont{and}
  \bibinfo{author}{\bibfnamefont{S.~U.} \bibnamefont{Egelhaaf}},
  \bibinfo{journal}{Soft Matter} \textbf{\bibinfo{volume}{18}},
  \bibinfo{pages}{4699} (\bibinfo{year}{2022}), ISSN \bibinfo{issn}{1744-683X}.

\bibitem[{\citenamefont{Kulkarni and Morris}(2009)}]{Kulkarni:2009un}
\bibinfo{author}{\bibfnamefont{S.~D.} \bibnamefont{Kulkarni}} \bibnamefont{and}
  \bibinfo{author}{\bibfnamefont{J.~F.} \bibnamefont{Morris}},
  \bibinfo{journal}{Journal of Rheology} \textbf{\bibinfo{volume}{53}},
  \bibinfo{pages}{417} (\bibinfo{year}{2009}), ISSN \bibinfo{issn}{0148-6055}.

\bibitem[{\citenamefont{Pieper and Schmid}(2016)}]{Pieper:2016vo}
\bibinfo{author}{\bibfnamefont{S.}~\bibnamefont{Pieper}} \bibnamefont{and}
  \bibinfo{author}{\bibfnamefont{H.-J.} \bibnamefont{Schmid}},
  \bibinfo{journal}{Journal of Non-Newtonian Fluid Mechanics}
  \textbf{\bibinfo{volume}{234}}, \bibinfo{pages}{1} (\bibinfo{year}{2016}),
  ISSN \bibinfo{issn}{0377-0257}.

\bibitem[{\citenamefont{K{\"u}{\c c}{\"u}ks{\"o}nmez and
  Servantie}(2020)}]{Kucuksonmez:2020vl}
\bibinfo{author}{\bibfnamefont{E.}~\bibnamefont{K{\"u}{\c c}{\"u}ks{\"o}nmez}}
  \bibnamefont{and}
  \bibinfo{author}{\bibfnamefont{J.}~\bibnamefont{Servantie}},
  \bibinfo{journal}{Physical Review E} \textbf{\bibinfo{volume}{102}},
  \bibinfo{pages}{012604} (\bibinfo{year}{2020}), ISSN
  \bibinfo{issn}{2470-0045}.

\bibitem[{\citenamefont{Myung et~al.}(2013)\citenamefont{Myung, Song, and
  Ahn}}]{Myung:2013uf}
\bibinfo{author}{\bibfnamefont{J.~S.} \bibnamefont{Myung}},
  \bibinfo{author}{\bibfnamefont{S.}~\bibnamefont{Song}}, \bibnamefont{and}
  \bibinfo{author}{\bibfnamefont{K.~H.} \bibnamefont{Ahn}},
  \bibinfo{journal}{Journal of Non-Newtonian Fluid Mechanics}
  \textbf{\bibinfo{volume}{199}}, \bibinfo{pages}{29} (\bibinfo{year}{2013}),
  ISSN \bibinfo{issn}{0377-0257}.

\bibitem[{\citenamefont{Morris}(2020)}]{Morris:2020uy}
\bibinfo{author}{\bibfnamefont{J.~F.} \bibnamefont{Morris}},
  \bibinfo{journal}{Annual Review of Fluid Mechanics}
  \textbf{\bibinfo{volume}{52}}, \bibinfo{pages}{121} (\bibinfo{year}{2020}),
  ISSN \bibinfo{issn}{0066-4189}.

\bibitem[{\citenamefont{Lee et~al.}(2018)\citenamefont{Lee, Jiang, Wang, Sandy,
  Narayanan, and Lin}}]{Lin:2018aa}
\bibinfo{author}{\bibfnamefont{J.}~\bibnamefont{Lee}},
  \bibinfo{author}{\bibfnamefont{Z.}~\bibnamefont{Jiang}},
  \bibinfo{author}{\bibfnamefont{J.}~\bibnamefont{Wang}},
  \bibinfo{author}{\bibfnamefont{A.~R.} \bibnamefont{Sandy}},
  \bibinfo{author}{\bibfnamefont{S.}~\bibnamefont{Narayanan}},
  \bibnamefont{and} \bibinfo{author}{\bibfnamefont{X.-M.} \bibnamefont{Lin}},
  \bibinfo{journal}{Physical Review Letters} \textbf{\bibinfo{volume}{120}}
  (\bibinfo{year}{2018}), ISSN \bibinfo{issn}{0031-9007}, \bibinfo{note}{not
  sure what this means}.

\bibitem[{\citenamefont{Goyal et~al.}(2022)\citenamefont{Goyal, Gado, Jones,
  and Martys}}]{Goyal:2022wa}
\bibinfo{author}{\bibfnamefont{A.}~\bibnamefont{Goyal}},
  \bibinfo{author}{\bibfnamefont{E.~D.} \bibnamefont{Gado}},
  \bibinfo{author}{\bibfnamefont{S.~Z.} \bibnamefont{Jones}}, \bibnamefont{and}
  \bibinfo{author}{\bibfnamefont{N.~S.} \bibnamefont{Martys}},
  \bibinfo{journal}{Journal of Rheology} \textbf{\bibinfo{volume}{66}},
  \bibinfo{pages}{1055} (\bibinfo{year}{2022}), ISSN \bibinfo{issn}{0148-6055}.

\bibitem[{\citenamefont{Boersma et~al.}(1991)\citenamefont{Boersma, Baets,
  Laven, and Stein}}]{Boersma:1991wi}
\bibinfo{author}{\bibfnamefont{W.~H.} \bibnamefont{Boersma}},
  \bibinfo{author}{\bibfnamefont{P.~J.~M.} \bibnamefont{Baets}},
  \bibinfo{author}{\bibfnamefont{J.}~\bibnamefont{Laven}}, \bibnamefont{and}
  \bibinfo{author}{\bibfnamefont{H.~N.} \bibnamefont{Stein}},
  \bibinfo{journal}{Journal of Rheology} \textbf{\bibinfo{volume}{35}},
  \bibinfo{pages}{1093} (\bibinfo{year}{1991}), ISSN \bibinfo{issn}{0148-6055}.

\bibitem[{\citenamefont{Lootens et~al.}(2003)\citenamefont{Lootens, Damme, and
  H{\'e}braud}}]{Lootens:2003ur}
\bibinfo{author}{\bibfnamefont{D.}~\bibnamefont{Lootens}},
  \bibinfo{author}{\bibfnamefont{H.}~\bibnamefont{Damme}}, \bibnamefont{and}
  \bibinfo{author}{\bibfnamefont{P.}~\bibnamefont{H{\'e}braud}},
  \bibinfo{journal}{Physical Review Letters} \textbf{\bibinfo{volume}{90}},
  \bibinfo{pages}{178301} (\bibinfo{year}{2003}), ISSN
  \bibinfo{issn}{1079-7114}.

\bibitem[{\citenamefont{Nakanishi et~al.}(2012)\citenamefont{Nakanishi,
  Nagahiro, and Mitarai}}]{Nakanishi:2012ux}
\bibinfo{author}{\bibfnamefont{H.}~\bibnamefont{Nakanishi}},
  \bibinfo{author}{\bibfnamefont{S.-i.} \bibnamefont{Nagahiro}},
  \bibnamefont{and} \bibinfo{author}{\bibfnamefont{N.}~\bibnamefont{Mitarai}},
  \bibinfo{journal}{Physical Review E} \textbf{\bibinfo{volume}{85}},
  \bibinfo{pages}{011401} (\bibinfo{year}{2012}), ISSN
  \bibinfo{issn}{1539-3755}.

\bibitem[{\citenamefont{Guy et~al.}(2015)\citenamefont{Guy, Hermes, and
  Poon}}]{Guy:2015wc}
\bibinfo{author}{\bibfnamefont{B.~M.} \bibnamefont{Guy}},
  \bibinfo{author}{\bibfnamefont{M.}~\bibnamefont{Hermes}}, \bibnamefont{and}
  \bibinfo{author}{\bibfnamefont{W.~C.~K.} \bibnamefont{Poon}},
  \bibinfo{journal}{Physical Review Letters} \textbf{\bibinfo{volume}{115}},
  \bibinfo{pages}{088304} (\bibinfo{year}{2015}), ISSN
  \bibinfo{issn}{0031-9007}.

\bibitem[{\citenamefont{Nagahiro and Nakanishi}(2016)}]{Nagahiro:2016wu}
\bibinfo{author}{\bibfnamefont{S.-I.} \bibnamefont{Nagahiro}} \bibnamefont{and}
  \bibinfo{author}{\bibfnamefont{H.}~\bibnamefont{Nakanishi}},
  \bibinfo{journal}{Physical review. E} \textbf{\bibinfo{volume}{94}},
  \bibinfo{pages}{062614} (\bibinfo{year}{2016}), ISSN
  \bibinfo{issn}{2470-0053}.

\bibitem[{\citenamefont{Hermes et~al.}(2016)\citenamefont{Hermes, Guy, Poon,
  Poy, Cates, and Wyart}}]{Hermes:2016wj}
\bibinfo{author}{\bibfnamefont{M.}~\bibnamefont{Hermes}},
  \bibinfo{author}{\bibfnamefont{B.~M.} \bibnamefont{Guy}},
  \bibinfo{author}{\bibfnamefont{W.~C.~K.} \bibnamefont{Poon}},
  \bibinfo{author}{\bibfnamefont{G.}~\bibnamefont{Poy}},
  \bibinfo{author}{\bibfnamefont{M.~E.} \bibnamefont{Cates}}, \bibnamefont{and}
  \bibinfo{author}{\bibfnamefont{M.}~\bibnamefont{Wyart}},
  \bibinfo{journal}{Journal of Rheology} \textbf{\bibinfo{volume}{60}},
  \bibinfo{pages}{905} (\bibinfo{year}{2016}), ISSN \bibinfo{issn}{0148-6055},
  \eprint{1511.08011}.

\bibitem[{\citenamefont{Rathee et~al.}(2017)\citenamefont{Rathee, Blair, and
  Urbach}}]{Rathee:2017ut}
\bibinfo{author}{\bibfnamefont{V.}~\bibnamefont{Rathee}},
  \bibinfo{author}{\bibfnamefont{D.~L.} \bibnamefont{Blair}}, \bibnamefont{and}
  \bibinfo{author}{\bibfnamefont{J.~S.} \bibnamefont{Urbach}},
  \bibinfo{journal}{Proceedings of the National Academy of Sciences}
  \textbf{\bibinfo{volume}{114}}, \bibinfo{pages}{8740} (\bibinfo{year}{2017}),
  ISSN \bibinfo{issn}{0027-8424}.

\bibitem[{\citenamefont{Saint-Michel et~al.}(2018)\citenamefont{Saint-Michel,
  Gibaud, and Manneville}}]{Saint-Michel:2018vw}
\bibinfo{author}{\bibfnamefont{B.}~\bibnamefont{Saint-Michel}},
  \bibinfo{author}{\bibfnamefont{T.}~\bibnamefont{Gibaud}}, \bibnamefont{and}
  \bibinfo{author}{\bibfnamefont{S.}~\bibnamefont{Manneville}},
  \bibinfo{journal}{Physical Review X} \textbf{\bibinfo{volume}{8}},
  \bibinfo{pages}{031006} (\bibinfo{year}{2018}), ISSN
  \bibinfo{issn}{2160-3308}.

\bibitem[{\citenamefont{Ovarlez et~al.}(2020)\citenamefont{Ovarlez, Le, Smit,
  Fall, Mari, Chatt{\'e}, and Colin}}]{Ovarlez:2020ul}
\bibinfo{author}{\bibfnamefont{G.}~\bibnamefont{Ovarlez}},
  \bibinfo{author}{\bibfnamefont{A.~V.~N.} \bibnamefont{Le}},
  \bibinfo{author}{\bibfnamefont{W.~J.} \bibnamefont{Smit}},
  \bibinfo{author}{\bibfnamefont{A.}~\bibnamefont{Fall}},
  \bibinfo{author}{\bibfnamefont{R.}~\bibnamefont{Mari}},
  \bibinfo{author}{\bibfnamefont{G.}~\bibnamefont{Chatt{\'e}}},
  \bibnamefont{and} \bibinfo{author}{\bibfnamefont{A.}~\bibnamefont{Colin}},
  \bibinfo{journal}{Science Advances} \textbf{\bibinfo{volume}{6}},
  \bibinfo{pages}{eaay5589} (\bibinfo{year}{2020}), ISSN
  \bibinfo{issn}{2375-2548}.

\bibitem[{\citenamefont{Gauthier et~al.}(2021)\citenamefont{Gauthier, Pruvost,
  Gamache, and Colin}}]{Gauthier:2021vg}
\bibinfo{author}{\bibfnamefont{A.}~\bibnamefont{Gauthier}},
  \bibinfo{author}{\bibfnamefont{M.}~\bibnamefont{Pruvost}},
  \bibinfo{author}{\bibfnamefont{O.}~\bibnamefont{Gamache}}, \bibnamefont{and}
  \bibinfo{author}{\bibfnamefont{A.}~\bibnamefont{Colin}},
  \bibinfo{journal}{Journal of Rheology} \textbf{\bibinfo{volume}{65}},
  \bibinfo{pages}{583} (\bibinfo{year}{2021}), ISSN \bibinfo{issn}{0148-6055}.

\bibitem[{\citenamefont{Maharjan et~al.}(2021)\citenamefont{Maharjan, O'Reilly,
  Postiglione, Klimenko, and Brown}}]{Maharjan:2021ud}
\bibinfo{author}{\bibfnamefont{R.}~\bibnamefont{Maharjan}},
  \bibinfo{author}{\bibfnamefont{E.}~\bibnamefont{O'Reilly}},
  \bibinfo{author}{\bibfnamefont{T.}~\bibnamefont{Postiglione}},
  \bibinfo{author}{\bibfnamefont{N.}~\bibnamefont{Klimenko}}, \bibnamefont{and}
  \bibinfo{author}{\bibfnamefont{E.}~\bibnamefont{Brown}},
  \bibinfo{journal}{Physical Review E} \textbf{\bibinfo{volume}{103}},
  \bibinfo{pages}{012603} (\bibinfo{year}{2021}), ISSN
  \bibinfo{issn}{2470-0045}, \eprint{2004.14316}.

\bibitem[{\citenamefont{Rathee et~al.}(2020{\natexlab{a}})\citenamefont{Rathee,
  Blair, and Urbach}}]{Rathee:2020un}
\bibinfo{author}{\bibfnamefont{V.}~\bibnamefont{Rathee}},
  \bibinfo{author}{\bibfnamefont{D.~L.} \bibnamefont{Blair}}, \bibnamefont{and}
  \bibinfo{author}{\bibfnamefont{J.~S.} \bibnamefont{Urbach}},
  \bibinfo{journal}{Soft Matter} \textbf{\bibinfo{volume}{17}},
  \bibinfo{pages}{1337} (\bibinfo{year}{2020}{\natexlab{a}}), ISSN
  \bibinfo{issn}{1744-683X}.

\bibitem[{\citenamefont{Rathee et~al.}(2020{\natexlab{b}})\citenamefont{Rathee,
  Blair, and Urbach}}]{Rathee:2020wi}
\bibinfo{author}{\bibfnamefont{V.}~\bibnamefont{Rathee}},
  \bibinfo{author}{\bibfnamefont{D.~L.} \bibnamefont{Blair}}, \bibnamefont{and}
  \bibinfo{author}{\bibfnamefont{J.~S.} \bibnamefont{Urbach}},
  \bibinfo{journal}{Journal of Rheology} \textbf{\bibinfo{volume}{64}},
  \bibinfo{pages}{299} (\bibinfo{year}{2020}{\natexlab{b}}), ISSN
  \bibinfo{issn}{0148-6055}.

\bibitem[{\citenamefont{Rathee et~al.}(2020{\natexlab{c}})\citenamefont{Rathee,
  Arora, Blair, Urbach, Sood, and Ganapathy}}]{Ganapathy:2020aa}
\bibinfo{author}{\bibfnamefont{V.}~\bibnamefont{Rathee}},
  \bibinfo{author}{\bibfnamefont{S.}~\bibnamefont{Arora}},
  \bibinfo{author}{\bibfnamefont{D.~L.} \bibnamefont{Blair}},
  \bibinfo{author}{\bibfnamefont{J.~S.} \bibnamefont{Urbach}},
  \bibinfo{author}{\bibfnamefont{A.~K.} \bibnamefont{Sood}}, \bibnamefont{and}
  \bibinfo{author}{\bibfnamefont{R.}~\bibnamefont{Ganapathy}},
  \bibinfo{journal}{Physical Review E} \textbf{\bibinfo{volume}{101}},
  \bibinfo{pages}{040601} (\bibinfo{year}{2020}{\natexlab{c}}), ISSN
  \bibinfo{issn}{2470-0045}, \eprint{1906.06356}.

\bibitem[{\citenamefont{Rathee et~al.}(2022)\citenamefont{Rathee, Miller,
  Blair, and Urbach}}]{Rathee:2022vy}
\bibinfo{author}{\bibfnamefont{V.}~\bibnamefont{Rathee}},
  \bibinfo{author}{\bibfnamefont{J.}~\bibnamefont{Miller}},
  \bibinfo{author}{\bibfnamefont{D.~L.} \bibnamefont{Blair}}, \bibnamefont{and}
  \bibinfo{author}{\bibfnamefont{J.~S.} \bibnamefont{Urbach}},
  \bibinfo{journal}{Proceedings of the National Academy of Sciences}
  \textbf{\bibinfo{volume}{119}}, \bibinfo{pages}{e2203795119}
  (\bibinfo{year}{2022}), ISSN \bibinfo{issn}{0027-8424}.

\bibitem[{\citenamefont{Dutta et~al.}(2013)\citenamefont{Dutta, Mbi, Arevalo,
  and Blair}}]{Dutta:2013aa}
\bibinfo{author}{\bibfnamefont{S.~K.} \bibnamefont{Dutta}},
  \bibinfo{author}{\bibfnamefont{A.}~\bibnamefont{Mbi}},
  \bibinfo{author}{\bibfnamefont{R.~C.} \bibnamefont{Arevalo}},
  \bibnamefont{and} \bibinfo{author}{\bibfnamefont{D.~L.} \bibnamefont{Blair}},
  \bibinfo{journal}{Rev Sci Instrum} \textbf{\bibinfo{volume}{84}},
  \bibinfo{pages}{063702} (\bibinfo{year}{2013}).

\bibitem[{\citenamefont{Besseling et~al.}(2009)\citenamefont{Besseling, Isa,
  Weeks, and Poon}}]{Poon:2009aa}
\bibinfo{author}{\bibfnamefont{R.}~\bibnamefont{Besseling}},
  \bibinfo{author}{\bibfnamefont{L.}~\bibnamefont{Isa}},
  \bibinfo{author}{\bibfnamefont{E.~R.} \bibnamefont{Weeks}}, \bibnamefont{and}
  \bibinfo{author}{\bibfnamefont{W.~C.} \bibnamefont{Poon}},
  \bibinfo{journal}{Advances in Colloid and Interface Science}
  \textbf{\bibinfo{volume}{146}}, \bibinfo{pages}{1} (\bibinfo{year}{2009}),
  ISSN \bibinfo{issn}{0001-8686}.

\bibitem[{\citenamefont{Taylor}(1938)}]{Taylor:1938ul}
\bibinfo{author}{\bibfnamefont{G.~I.} \bibnamefont{Taylor}},
  \bibinfo{journal}{Proceedings of the Royal Society of London. Series A -
  Mathematical and Physical Sciences} \textbf{\bibinfo{volume}{164}},
  \bibinfo{pages}{476} (\bibinfo{year}{1938}), ISSN \bibinfo{issn}{0080-4630}.

\bibitem[{\citenamefont{He et~al.}(2017)\citenamefont{He, Jin, and
  Yang}}]{He:2017ug}
\bibinfo{author}{\bibfnamefont{G.}~\bibnamefont{He}},
  \bibinfo{author}{\bibfnamefont{G.}~\bibnamefont{Jin}}, \bibnamefont{and}
  \bibinfo{author}{\bibfnamefont{Y.}~\bibnamefont{Yang}},
  \bibinfo{journal}{Annual Review of Fluid Mechanics}
  \textbf{\bibinfo{volume}{49}}, \bibinfo{pages}{51} (\bibinfo{year}{2017}),
  ISSN \bibinfo{issn}{0066-4189}.

\bibitem[{\citenamefont{Koynov and Butt}(2012)}]{koynov:212aa}
\bibinfo{author}{\bibfnamefont{K.}~\bibnamefont{Koynov}} \bibnamefont{and}
  \bibinfo{author}{\bibfnamefont{H.-J.} \bibnamefont{Butt}},
  \bibinfo{journal}{Current Opinion in Colloid \& Interface Science}
  \textbf{\bibinfo{volume}{17}}, \bibinfo{pages}{377} (\bibinfo{year}{2012}),
  ISSN \bibinfo{issn}{1359-0294}.

\bibitem[{\citenamefont{Dong and Ren}(2014)}]{Dong:2014aa}
\bibinfo{author}{\bibfnamefont{C.}~\bibnamefont{Dong}} \bibnamefont{and}
  \bibinfo{author}{\bibfnamefont{J.}~\bibnamefont{Ren}},
  \bibinfo{journal}{ELECTROPHORESIS} \textbf{\bibinfo{volume}{35}},
  \bibinfo{pages}{2267} (\bibinfo{year}{2014}), ISSN \bibinfo{issn}{0173-0835}.

\bibitem[{\citenamefont{Kunst et~al.}(2002)\citenamefont{Kunst, Schots, and
  Visser}}]{kunst:2002aa}
\bibinfo{author}{\bibfnamefont{B.~H.} \bibnamefont{Kunst}},
  \bibinfo{author}{\bibfnamefont{A.}~\bibnamefont{Schots}}, \bibnamefont{and}
  \bibinfo{author}{\bibfnamefont{A.~J. W.~G.} \bibnamefont{Visser}},
  \bibinfo{journal}{Analytical Chemistry} \textbf{\bibinfo{volume}{74}},
  \bibinfo{pages}{5350} (\bibinfo{year}{2002}), ISSN \bibinfo{issn}{0003-2700}.

\bibitem[{\citenamefont{Gösch et~al.}(2000)\citenamefont{Gösch, Blom, Holm,
  Heino, and Rigler}}]{Gosch:2000aa}
\bibinfo{author}{\bibfnamefont{M.}~\bibnamefont{Gösch}},
  \bibinfo{author}{\bibfnamefont{H.}~\bibnamefont{Blom}},
  \bibinfo{author}{\bibfnamefont{J.}~\bibnamefont{Holm}},
  \bibinfo{author}{\bibfnamefont{T.}~\bibnamefont{Heino}}, \bibnamefont{and}
  \bibinfo{author}{\bibfnamefont{R.}~\bibnamefont{Rigler}},
  \bibinfo{journal}{Analytical Chemistry} \textbf{\bibinfo{volume}{72}},
  \bibinfo{pages}{3260} (\bibinfo{year}{2000}), ISSN \bibinfo{issn}{0003-2700}.

\bibitem[{\citenamefont{Pan et~al.}(2007)\citenamefont{Pan, Yu, Shi, Korzh, and
  Wohland}}]{pan:2007aa}
\bibinfo{author}{\bibfnamefont{X.}~\bibnamefont{Pan}},
  \bibinfo{author}{\bibfnamefont{H.}~\bibnamefont{Yu}},
  \bibinfo{author}{\bibfnamefont{X.}~\bibnamefont{Shi}},
  \bibinfo{author}{\bibfnamefont{V.}~\bibnamefont{Korzh}}, \bibnamefont{and}
  \bibinfo{author}{\bibfnamefont{T.}~\bibnamefont{Wohland}},
  \bibinfo{journal}{Journal of Biomedical Optics}
  \textbf{\bibinfo{volume}{12}}, \bibinfo{pages}{014034}
  (\bibinfo{year}{2007}), ISSN \bibinfo{issn}{1083-3668}.

\bibitem[{\citenamefont{Pan et~al.}(2009)\citenamefont{Pan, Shi, Korzh, Yu, and
  Wohland}}]{pan:2009aa}
\bibinfo{author}{\bibfnamefont{X.}~\bibnamefont{Pan}},
  \bibinfo{author}{\bibfnamefont{X.}~\bibnamefont{Shi}},
  \bibinfo{author}{\bibfnamefont{V.}~\bibnamefont{Korzh}},
  \bibinfo{author}{\bibfnamefont{H.}~\bibnamefont{Yu}}, \bibnamefont{and}
  \bibinfo{author}{\bibfnamefont{T.}~\bibnamefont{Wohland}},
  \bibinfo{journal}{Journal of Biomedical Optics}
  \textbf{\bibinfo{volume}{14}}, \bibinfo{pages}{024049}
  (\bibinfo{year}{2009}), ISSN \bibinfo{issn}{1083-3668}.

\bibitem[{\citenamefont{Ackerson}(1990)}]{Ackerson:1990uh}
\bibinfo{author}{\bibfnamefont{B.~J.} \bibnamefont{Ackerson}},
  \bibinfo{journal}{Journal of Rheology} \textbf{\bibinfo{volume}{34}},
  \bibinfo{pages}{553} (\bibinfo{year}{1990}), ISSN \bibinfo{issn}{0148-6055}.

\bibitem[{\citenamefont{Meer}(2016)}]{devaraj:2916aa}
\bibinfo{author}{\bibfnamefont{D.~v.~d.} \bibnamefont{Meer}},
  \bibinfo{journal}{Annual Review of Fluid Mechanics}
  \textbf{\bibinfo{volume}{49}}, \bibinfo{pages}{1} (\bibinfo{year}{2016}),
  ISSN \bibinfo{issn}{0066-4189}.

\bibitem[{\citenamefont{O'Neill et~al.}(2019)\citenamefont{O'Neill, Royer, and
  Poon}}]{ONeill:2019tv}
\bibinfo{author}{\bibfnamefont{R.~E.} \bibnamefont{O'Neill}},
  \bibinfo{author}{\bibfnamefont{J.~R.} \bibnamefont{Royer}}, \bibnamefont{and}
  \bibinfo{author}{\bibfnamefont{W.~C.~K.} \bibnamefont{Poon}},
  \bibinfo{journal}{Physical Review Letters} \textbf{\bibinfo{volume}{123}},
  \bibinfo{pages}{128002} (\bibinfo{year}{2019}), ISSN
  \bibinfo{issn}{0031-9007}, \eprint{1808.09950}.

\bibitem[{\citenamefont{Style et~al.}(2014)\citenamefont{Style, Boltyanskiy,
  German, Hyland, MacMinn, Mertz, Wilen, Xu, and Dufresne}}]{Style:2014aa}
\bibinfo{author}{\bibfnamefont{R.~W.} \bibnamefont{Style}},
  \bibinfo{author}{\bibfnamefont{R.}~\bibnamefont{Boltyanskiy}},
  \bibinfo{author}{\bibfnamefont{G.~K.} \bibnamefont{German}},
  \bibinfo{author}{\bibfnamefont{C.}~\bibnamefont{Hyland}},
  \bibinfo{author}{\bibfnamefont{C.~W.} \bibnamefont{MacMinn}},
  \bibinfo{author}{\bibfnamefont{A.~F.} \bibnamefont{Mertz}},
  \bibinfo{author}{\bibfnamefont{L.~A.} \bibnamefont{Wilen}},
  \bibinfo{author}{\bibfnamefont{Y.}~\bibnamefont{Xu}}, \bibnamefont{and}
  \bibinfo{author}{\bibfnamefont{E.~R.} \bibnamefont{Dufresne}},
  \bibinfo{journal}{Soft Matter} \textbf{\bibinfo{volume}{10}},
  \bibinfo{pages}{4047} (\bibinfo{year}{2014}), ISSN \bibinfo{issn}{1744-683X}.

\end{thebibliography}

\end{document}


\begin{center}

\textbf{Supplementary Figures \\  Order and density fluctuations near the boundary in sheared dense suspensions}\\
{Joia M. Miller, Daniel L. Blair, Jeffrey S. Urbach}
\end{center}



\begin{figure*}[h!]
\includegraphics[width=0.75\textwidth]{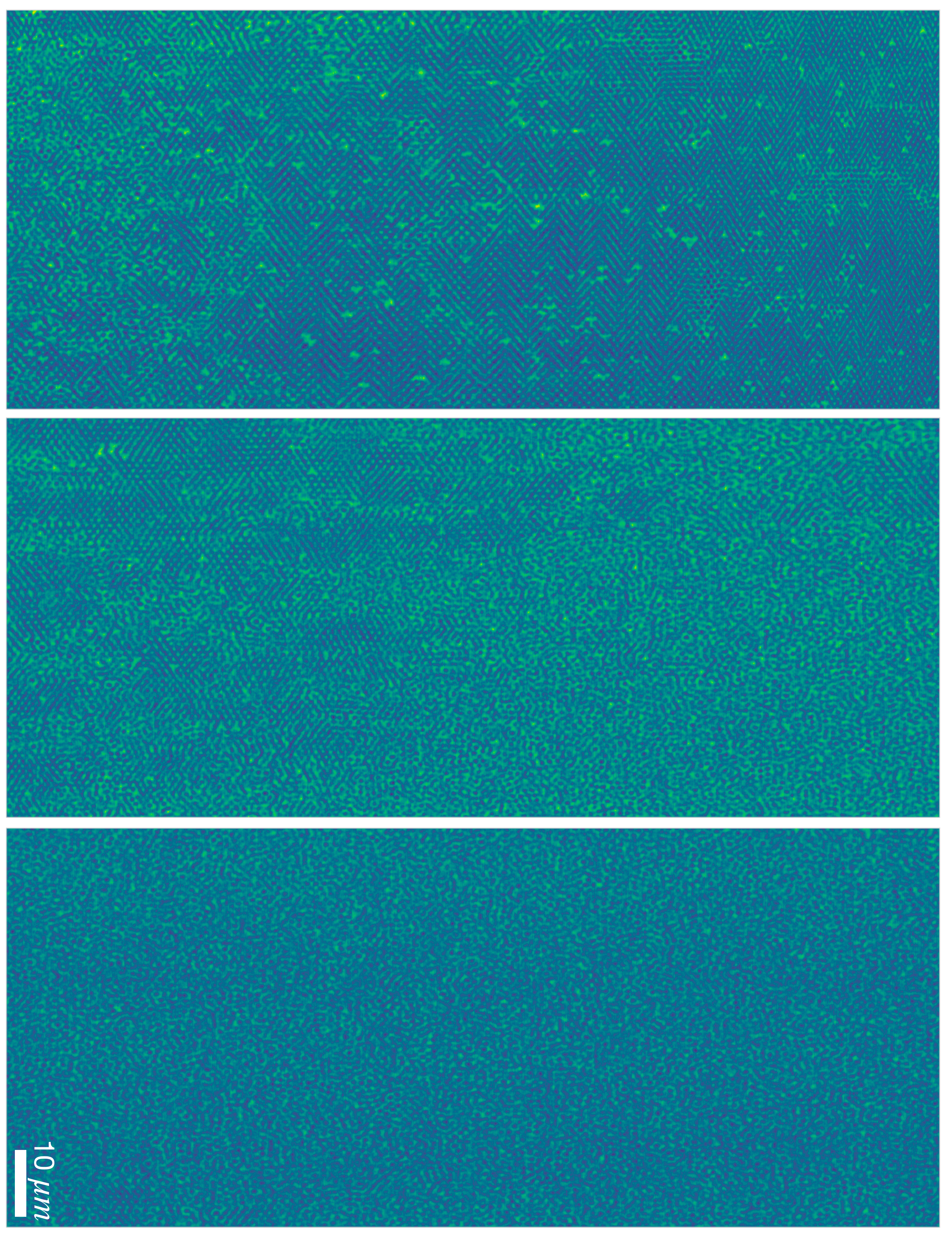}
\caption{ 
Rescaled images from kymographs shown in Fig. 5, with the spacing in the time (horizontal) direction scaled by the average speed in each individual set (512 scan lines). (20 Pa constant applied stress at heights of 3$\mu m$ (top), 10 $\mu m$ (middle), and 20 $\mu m$ above the bottom surface, $\phi \approx 0.55$). The bottom image covers the the same time range as Fig. 5, while the sets taken at slower speeds cover longer times so that the distance elapsed in the flow direction is approximately the same.)  }
\label{20Pa-images}
\end{figure*}

\begin{figure*}
\includegraphics[width=1\textwidth]{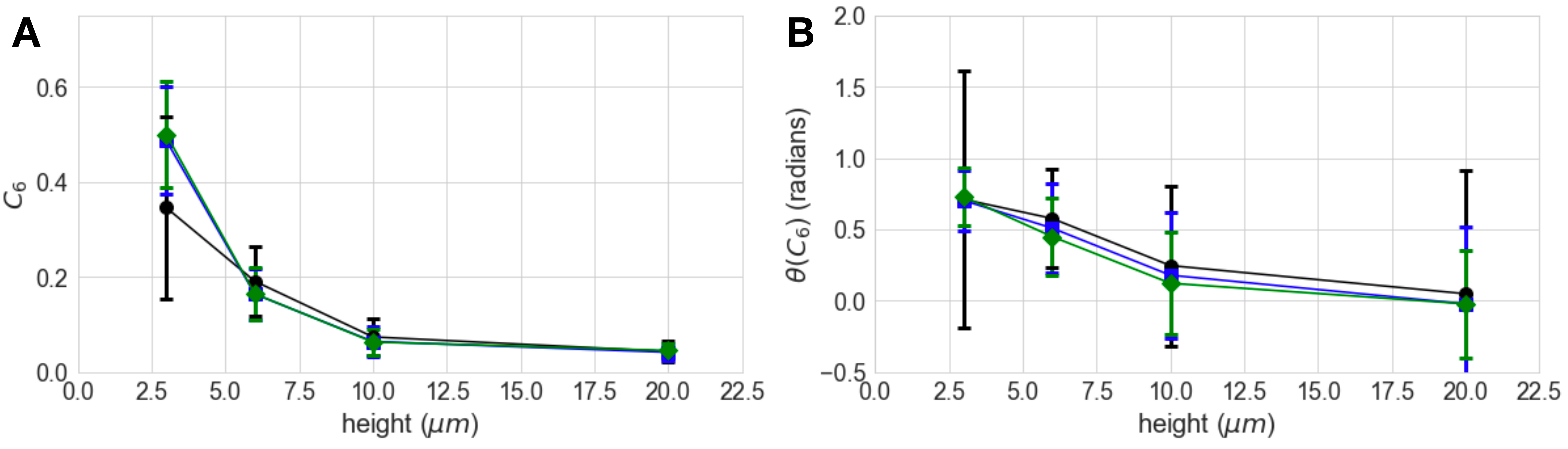}
\caption{A) Average orientational order parameter vs. height for constant shear rate, $\phi=0.52$.  B) Average angle of hexagonal order relative to flow direction. The value at $z=3 \mu m$ is approximately $\pi/6$.  By contrast, $\theta(C_6)\approx 0$ for all conditions at $\phi=0.56$, which corresponds to one of the hexagonal lattice vectors aligned with the flow axis. 
$10s^{-1}$
({\color{black}$\CIRCLE$}), 
$20s^{-1}$  
({\color{blue}$\blacksquare$ }), 
 $30s^{-1}$ 
 ({\color{green}$\blacklozenge$})}.
\label{SR_order}
\end{figure*}

\begin{figure*}
\includegraphics[width=1\textwidth]{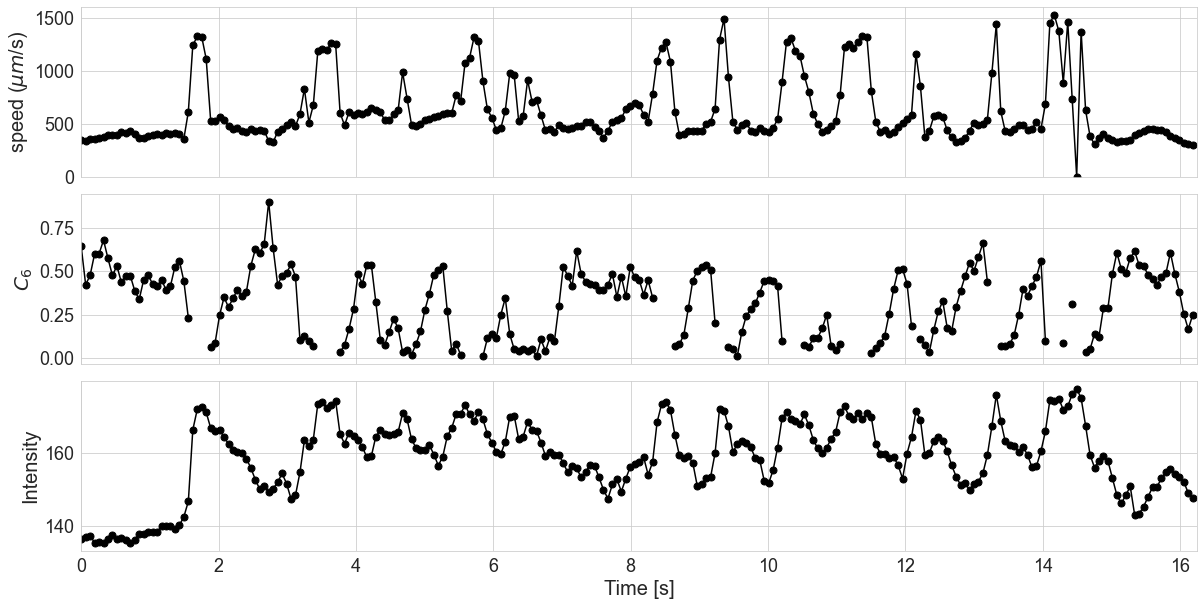}
\caption{Section of time series of speed, hexagonal order, and intensity generated from kymographs at applied stress 100 Pa and height 3 $\mu m$ during cluster of high speed events. The missing values  of $C_6$ arise because the algorithm used fails at very high speeds.  Visual inspection shows no discernible order in those kymographs. }
\label{100Pa_3um_cluster}
\end{figure*}

\begin{figure*}
\includegraphics[width=1\textwidth]{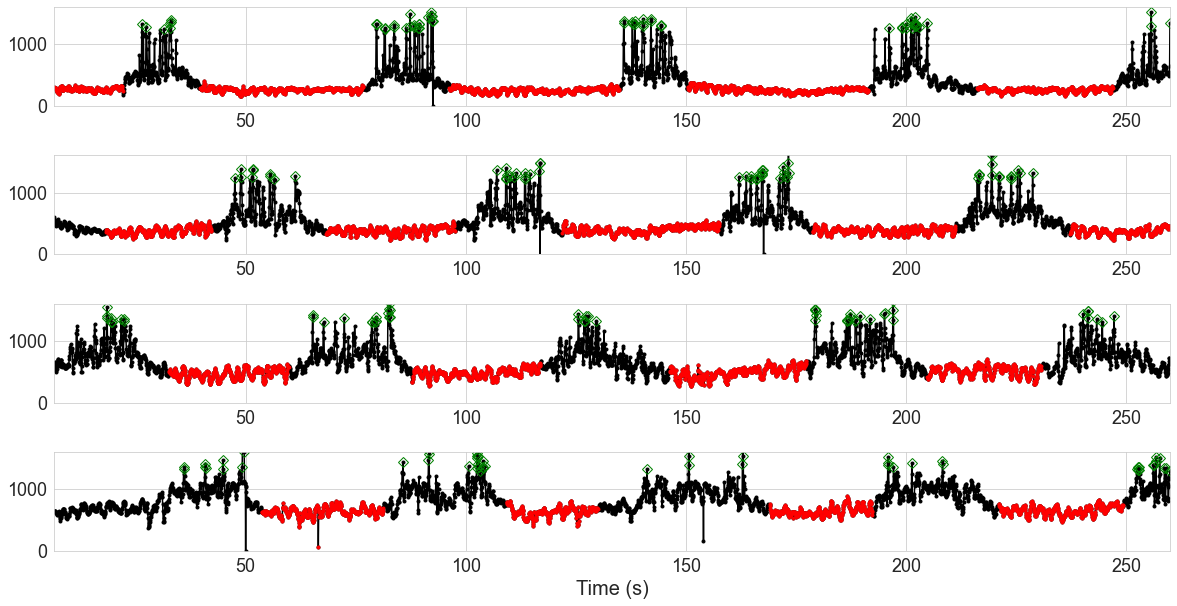}
\caption{Timeseries of speeds for 100 Pa applied stress at heights of 3, 6, 10, and 20 $\mu m$ with sections used for average speeds away from events (red) and at the 60 highest speeds for each height ({\color{green}$\Diamond$}) used for the averages shown in Fig. 12.}
\label{100Pa_segmentation}
\end{figure*}


